\documentclass[%
reprint,
 amsmath,amssymb,
aps,
]{revtex4-2}

\usepackage{graphicx}
\usepackage{dcolumn}
\usepackage{bm}
\usepackage[normalem]{ulem}
\usepackage{hyperref}

\usepackage{physics}
\usepackage{xcolor}

\begin{document}


\title{Emission and detection of ultrahigh frequency gravitational waves from highly eccentric orbits of compact binary systems}

\author{Pierre Jamet}
 \email{jamet@lpsc.in2p3.fr}
\author{Aurélien Barrau}
 \email{barrau@lpsc.in2p3.fr}
\author{Killian Martineau}
 \email{martineau@lpsc.in2p3.fr}
\affiliation{
 Laboratoire de Physique Subatomique et de Cosmologie, Univ. Grenoble Alpes, CNRS/IN2P3\\ 53 avenue des Martyrs, 38026 Grenoble Cedex, France
}


\date{\today}

\begin{abstract}
The ultrahigh frequency emission of gravitational waves by binary systems of black holes has recently been investigated in details in the framework of new experimental ideas around resonant cavities. In this article, we consider the case of elliptic trajectories. At fixed masses and frequency, we conclude that the total amount of energy radiated by the system within the bandwidth of the detector can be significantly higher than for circular orbits. However, owing to subtle experimental effects, the signal-to-noise ratio is, overall, a decreasing function of the eccentricity. Limits on the maximum distance at which a merging system of black holes can be detected derived are therefore {\it not} improved by considering elliptic trajectories, when compared to the circular case. The article is written as pedagogically as possible so as to be accessible to the nonfamiliar reader and possibly useful beyond the ultrahigh frequency case.
\end{abstract}

\maketitle

\section{\label{sec:1}Introduction}

Not long ago, gravitational waves were observed by the LIGO-Virgo Collaboration in the [$10 - 10^{4}$] Hz range \cite{KAGRA:2021vkt}. At lower frequencies, the LISA free-falling satellites should detect gravitational waves around [$10^{-4} – 1$] Hz \cite{amaroseoane2017laser}. Below, time delays in pulsar signals were measured by the international pulsar timing array consortium showing evidence for a stochastic background in the nHz range \cite{NANOGrav:2023gor, EPTA:2023fyk, Reardon:2023gzh}. On the other hand, gravitational waves could also be detected at very high frequency, say above the MHz, in the near future \cite{Goryachev:2021zzn, PhysRevD.90.102005, Berlin:2021txa, Berlin:2023grv, Domcke:2022rgu, Domcke:2020yzq, PhysRevD.104.023524}. This is the focus of the present article. A review of the candidate sources can be found in \cite{Aggarwal:2020olq}, and details about the expected signal are given in \cite{Berlin:2021txa}. The search for light black holes was considered in \cite{Herman:2020wao,Franciolini:2022htd}.\\

Recently, specific investigations of GHz emission from black holes on circular orbits were performed in \cite{Franciolini:2022htd, barrau2023} whereas hyperbolic trajectories were considered in \cite{teuscher2024,barrau2023prospect}. Quite naturally, to fully cover the topic we now focus on elliptic orbits for binary systems of black holes. The question that should be answered is  basically the following: for given masses, could an elliptic orbit be more favorable for detection than a circular one? Otherwise stated, we try to understand if results obtained for circular orbits can be improved by considering more general trajectories or if known results can be considered as optimistic estimates (therefore making the derived upper limits on the reachable distances conservative and reliable). It should be immediately emphasized that the answer is deeply nontrivial and depends crucially on the way the signal is detected. 

A key point to keep in mind is that GHz experiments have a very narrow bandwidth. This is inherent to the functioning of resonant cavities—with quality factors typically of the order of $10^5$. Although we carefully take into account its precise value and influence in the following, most of the intuition should be built on a reasoning at nearly fixed frequency. This is precisely why the comparison is subtle: we compare trajectories with different eccentricities but at fixed masses and for a fixed emission frequency.\\

When dealing with circular orbits, it is possible to consider, at the lowest order, that the evolution of the frequency of the emitted gravitational waves is {\it entirely} due to the emission itself. This is the methodology used in \cite{barrau2023}. The Newtonian dynamics would lead to a purely monochromatic signal without any evolution at all. The frequency drift is therefore only caused by the energy lost by gravitational radiation. On the other hand, for hyperbolic trajectories, it is possible to assume, as in \cite{barrau2023prospect}, that the backreaction is negligible at the lowest order and that the evolution of the frequency is {\it entirely} due to the variation of the time derivative of the position angle along the Newtonian trajectory. Otherwise stated, in this latter case and at this level of approximation, the evolution of the signal is fully governed by the highly nonperiodic path leading to a fast-varying instantaneous frequency without taking into account the energy lost by gravitational radiation. 

In the elliptic case, the situation is more involved, and one has to take into account both effects simultaneously: on a given orbit of fixed eccentricity the gravitational wave frequency varies very substantially with time (even fully ignoring backreaction) but, in addition, the eccentricity is also strongly time dependent because of the emission of gravitational waves. As we shall show in the following, this makes the situation quite complicated. We emphasize the aim of this study is not to give a definitive answer on the topic but to provide a clarification at the lowest nontrivial order. The naive expectation that eccentric orbits are easier to detect because the power emitted as gravitational waves gets an extra factor $F(e)=(1-e^2)^{-7/2}(1+73/24\,e^2 + 37/96\,e^4)$ per period is actually not correct.\\

In a nutshell, the renewed interest for gravitational waves in the GHz band is due to the understanding that resonant cavities located at the core of haloscope experiments (initially designed to search for axions), can be used as efficient gravitational wave detectors at very high frequencies ; see, \textit{e.g}, \cite{Berlin:2021txa, Valero:2024ncz}. The case of haloscopes operating at lower frequencies, typically in the [0.1 - 100 ] MHz range, is also under investigation \cite{Domcke:2022rgu, Domcke:2023bat}.
To set orders of magnitude, if one considers equal mass black holes and requires the gravitational wave frequency to be in the GHz band at the merging, the mass should be of the order of $10^{-6}$ $M\textsubscript{\(\odot\)}$. Obviously, only black holes of primordial origin \cite{Carr:2020xqk} can exist at such small masses. Very importantly, this should be taken as an upper bound and, in no way, as an estimate of the accessible masses. A system with smaller masses will simply be seen in the bandwidth of the instrument earlier in the inspiral process. It is mandatory to consider all possibilities as we do not know the actual masses of existing black holes (if any do exist) in this range. There is no reason for the real system be be tuned for the optimum experimental sensitivity. In addition, it was shown in \cite{barrau2023} that, for a wide range of masses, the smaller strain generated by smaller masses is compensated by the longer time spent in the bandwidth. Not to mention that, for masses close to saturating the upper bound imposed by the detection frequency, the formulas used in this work -- and in the previous studies -- are actually not reliable anymore. The reason is that the trajectory is no longer a conic.\\

In the following, we first explain the general parametrization used to describe the orbit and the characteristics of the emitted gravitational waves. We then present the results of numerical simulations for large masses so as to help the intuitive understanding of the situation. The main conclusions are then derived, focusing on the more physical case of smaller masses. We conclude with the limits of the approach and possible improvements.\\

\section{\label{sec:2}General parametrization and equations for elliptic orbits}


The figure in the Appendix displays the parametrization chosen for elliptic trajectories. 
The most obvious description is based on the semimajor and semiminor axes $a$ and $b$. It is also convenient to rely on the eccentricity $e = \sqrt{1 - b^2/a^2}$ instead of one of the axes. This is especially interesting for this study as it allows an easy and intuitive understanding of the trajectory as a deformation of the usual circular orbit. 
To keep in line with previous works and to emphasize variables of explicit interest, we also replace the other axis by the angular frequency at periapsis $\omega_p$. This happens to be very meaningful  because that particular frequency is the one corresponding to the maximum of the signal Fourier transform when the gravitational wave emission burst occurs. In order to efficiently define this parameter, we however need to add some physics to the mathematics of conics.\\


For two black holes of masses $m_1$ and $m_2$, we define the total mass $M = m_1 + m_2$, and the  reduced mass is $\mu = \frac{m_1 m_2}{M}$. With $\kappa \equiv \mu G$ and $G$ the gravitational constant, the angular frequency at periapsis reads as
\begin{equation}
  \omega_p = \sqrt{\frac{\kappa(1+e)}{a^3(1-e)^3}}.
\end{equation}

To fully describe the dynamics, we also need to introduce the specific position of the object of mass $\mu$ along the orbit. We choose to use the so-called true anomaly $\varphi$, as well as the instantaneous angular velocity $\omega$, obtained from Newtonian orbital mechanics:
\begin{equation}
  \omega = \omega_p\pqty{\frac{1 + e \cos\varphi}{1+e}}^2.
  \label{eq:puls}
\end{equation}
As an alternative useful parameter, one can also consider the distance to the focus: 
\begin{equation}
  r = \pqty{\frac{\kappa}{\omega_p^2}}^\frac{1}{3}\frac{1 + e}{1 + e \cos\varphi}.
  \label{eq:rad}
\end{equation}

The leading order derivation of the strain generated by this system is fully textbook \cite{maggiore2008} and some steps (with important remarks for the following) are given in the Appendix of this article. 
The resulting expressions for the strains are
\begin{widetext}
\begin{align}
  h_+ &= -\frac{\mu G}{R c^4}\pqty{\frac{\kappa\omega_p}{(1+e)^2}}^{\frac{2}{3}}\pqty{2e^2 + 5 e \cos\varphi + 4\cos 2\varphi + e \cos 3\varphi},\label{eq:hplus}\\
  h_\times &= -\frac{\mu G}{R c^4}\pqty{\frac{\kappa\omega_p}{(1+e)^2}}^{\frac{2}{3}} \sin\varphi\pqty{6 e + 8\cos\varphi + 2e\cos 2\varphi}.\label{eq:hcross}
\end{align}
\end{widetext}


The energy and angular momentum carried away by the emitted gravitational waves will obviously backreact on the dynamics of the source, eventually leading to the coalescence of the system. On top of the well-known decrease of the radius of the orbit, 
the eccentricity of elliptic trajectories also decreases, and this variation is usually  faster than that of the radius—so that the system first circularizes and then merges \cite{Maggiore:2007ulw}.

It is important to stress that the situation is tricky, even in this simple setting. Not only does the instantaneous frequency of emitted gravitational waves strongly vary along the orbit but, in addition, the parameters of the orbit evolve themselves in a nontrivial way. The main steps are summarized in the Appendix. 

The resulting differential equations are
\begin{align}
  \dot{a} &= -\frac{\mu G \kappa^2}{15 c^5}\frac{1}{a^3}\frac{1}{\pqty{1 - e^2}^{\frac{7}{2}}}\pqty{192 + 584 e^2 + 74 e^4},\label{eq:adiff}\\
  \dot{e} &= -\frac{\mu G \kappa^2}{15 c^5}\frac{1}{a^4}\frac{e}{\pqty{1 - e^2}^{\frac{5}{2}}}\pqty{304 + 121 e^2}.\label{eq:ediff}
\end{align}
Since the quantities of interest, such as the frequency of the signal as well as the strain, are written in terms of the angle $\varphi$, which is itself a complicated function of time, it is required to solve for this variable with a third differential equation (using the angular frequency):
\begin{equation}
  \dot{\varphi} = \omega(\varphi) = \sqrt{\frac{\kappa(1+e)}{a^3(1-e)^3}}\pqty{\frac{1 + e \cos\varphi}{1+e}}^2,
\end{equation}
where the time dependence of $e$ and $a$ are given by the previous equations.

\section{\label{sec:3}Numerical results}
\subsection{\label{sec:3_1}Numerical parameters and integration procedure}

Solving this differential system numerically is not straightforward. Three different characteristic timescales enter the dynamics. The first two, mentioned previously, are related to the frequency of the gravitational waves and to the orbital period, whereas the third one is the time to coalescence. They are obviously related to one another but can take widely different values spanning many orders of magnitude. 
For instance, in the case of a highly eccentric orbit with $e = 0.9$, the angular frequency at periapsis $\omega_p$ is nearly 50 times larger than the orbital frequency $\omega_0$.
In general, this ratio depends neither on the masses involved nor on the value of $\omega_p$, and is given by $\omega_p/\omega_0 = \sqrt{\frac{1+e}{(1-e)^3}}$.
We shall focus on the parameters leading to the best and clearest visualization—corresponding to the ones for which the time scales are comparable. We shall argue that the qualitative conclusions drawn from the specific examples presented here should hold in general.\\

For now, we choose for the masses $m_1 = m_2 = 1.5\times10^{-6}~M_\odot$, which correspond to primordial black holes that are unaffected by the Hawking evaporation (the specific question of the competitive effects between gravitational radiation and mass variation for two-body systems in circular orbits was considered in \cite{Blachier:2023ygh}). This makes the understanding of the physical behavior easier. From Eqs. (\ref{eq:hplus}) and (\ref{eq:hcross}), it is straightforward to have an idea of how the system would behave for asymmetrical values of the individual masses at fixed total mass (as the Newtonian orbit is fully determined by $M$). One should keep in mind that $m_1=m_2$ maximizes $\mu$ for a given $M$. Otherwise stated, the strain is expected to decrease for $m_1\neq m_2$, hence the SNR. We show examples, for smaller masses, later in the article.\\ 

For initial conditions, we first set the angular frequency at the periapsis, as it should obviously be somehow close to the frequency of the resonant cavity. Some freedom still remains in this choice since the average frequency of the signal shifts upward in time whereas, within the orbit, the instantaneous frequency, defined as $\dot{\varphi}/(2\pi)$ can be significantly smaller.  For simplicity, we always choose a starting frequency slightly under the lower end of the detector's bandwidth (typically one full bandwidth below) since only a negligible amount of energy (corresponding to hypereccentric trajectories considered very far away from the periapsis) can, this way, be lost in the calculation. It should anyway be stressed that this is a purely technical issue which does not impact the physical results. With $\omega_\mathrm{det.}$ the angular frequency to which the detector is sensitive (that is $\omega_\mathrm{det.}=2\pi\times 10^9$~rad.s$^{-1}$  for a frequency of $1$~GHz) and $Q$ the quality factor of the cavity (typically $Q=10^5$ in cases of relevance for this study), one can write the initial condition as $\omega_{p,0} = \omega_\mathrm{det.}\pqty{1-Q^{-1}}$. As in \cite{barrau2023,barrau2023prospect}, we choose the GrAHal experiment \cite{Grenet:2021vbb, GrenetFIPs22} as a benchmark but the conclusions remain true for all detectors based on resonant cavities operating around the GHz.  \\

The eccentricity is the main focus of this study—our goal is to understand its effect on a possible detection of the signal. It the next section, it will be varied over a wide range of values. To get an intuition of what is going on, we set its initial value at $e_0 = 0.9$, which helps underlying specific features arising from ellipses. At the most fundamental level, the very definition of the eccentricity is in itself subtle in general relativity (see \cite{Boschini:2024scu} and references therein for a recent review). We shall, however, be concerned here with the simpler—but still important—problem of properly defining, at the Newtonian level, what we mean by ``the eccentricity of the orbit" while this parameter is continuously (and, substantially, in the regime of interest) varying in time. From now on, we mostly use $e_1$, defined as the eccentricity of the orbit when the instantaneous frequency first enters the bandwidth.

The initial angle is also a parameter which has to be fixed. It turns out that, in the regime we consider, it is far from being a detail. For now, we will set it at the apoapsis, that is $\varphi_0 = -\pi$. For most of this study, that is for small masses, the numerical value chosen for $\varphi_0$ does not play a significant role, hence making the results both reliable and consistent (as the formulas used require anyway that the mass bound—fixed by the detector frequency—should not be saturated). For the ``academic" case of large masses considered at this stage, the effect of $\varphi_0$ can obviously become non-negligible. In this regime, the very concept of studying physical quantities as a function of eccentricity is however intrinsically ill-defined. We investigate in detail this effect for relevant masses in Sec. III C.\\

\begin{figure}
  \centering
  \includegraphics[width=.9\linewidth]{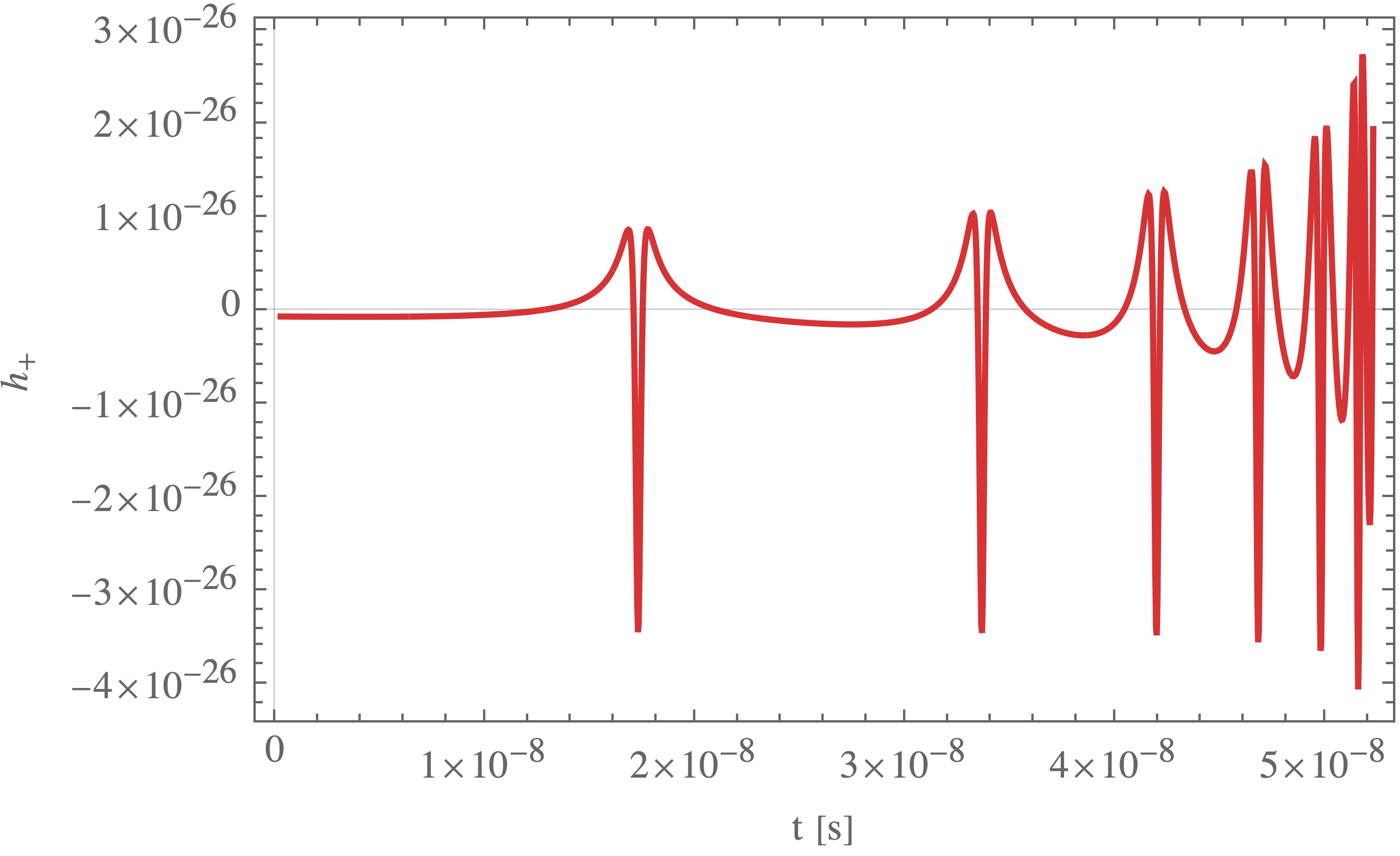}
  \caption{Time evolution of the ``plus'' polarization of the strain at a distance $R = 1$~Mpc from the source.}
  \label{fig:strain_plus}
\end{figure}
\begin{figure}
  \centering
  \includegraphics[width=.9\linewidth]{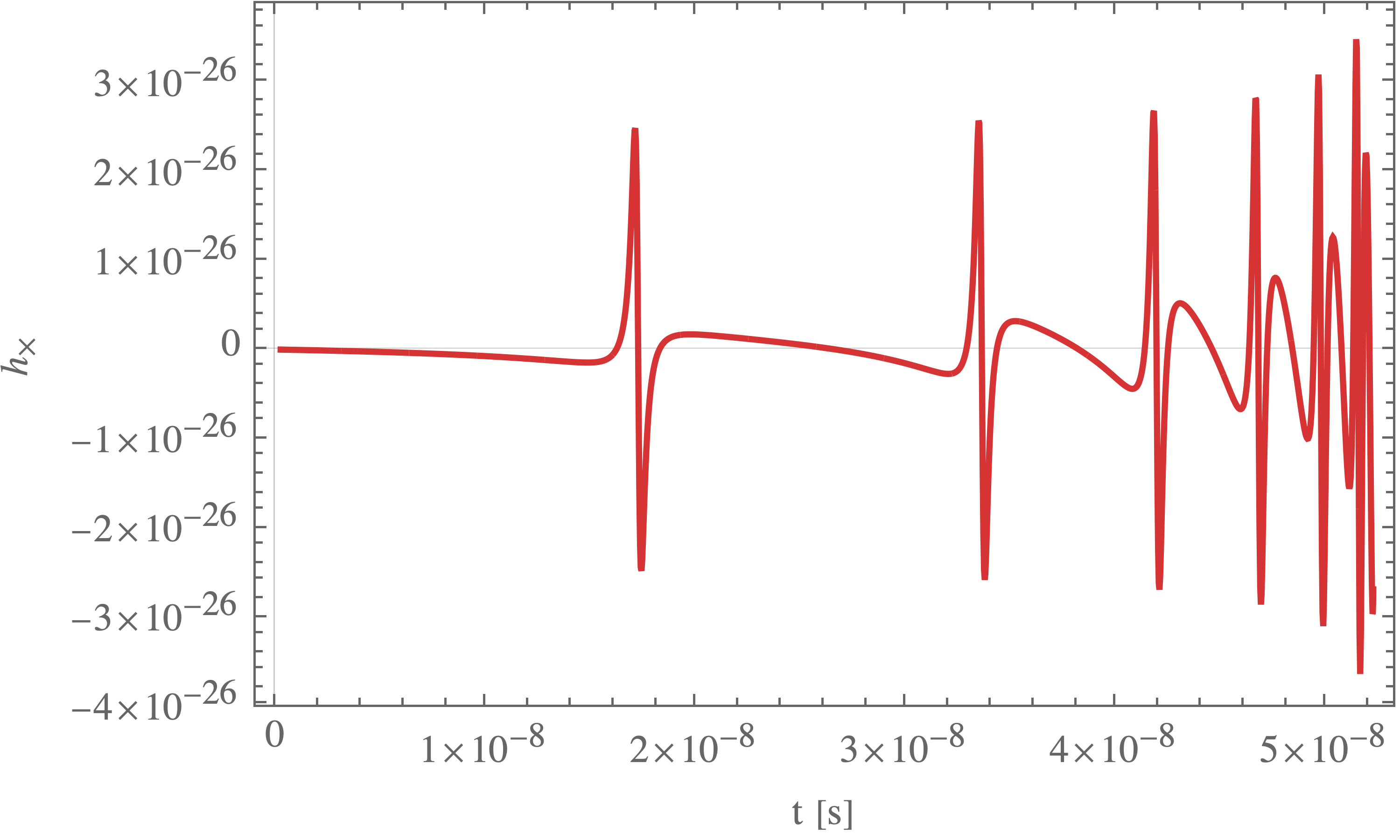}
  \caption{Time evolution of the ``cross'' polarization of the strain at a distance $R = 1$~Mpc from the source.}
  \label{fig:strain_cross}
\end{figure}

\begin{figure}
  \centering
  \includegraphics[width=.9\linewidth]{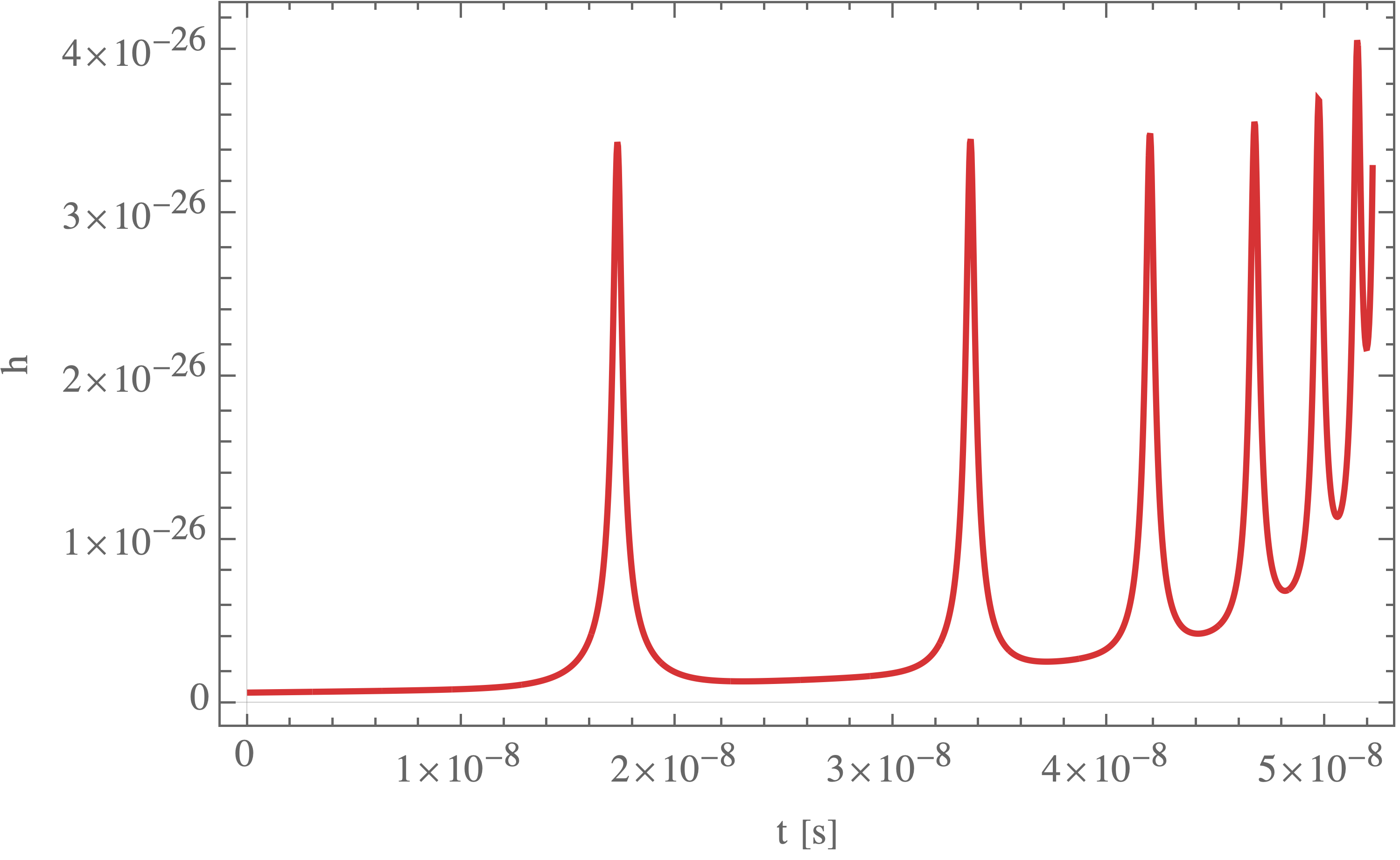}
  \caption{Time evolution of the total strain at a distance $R = 1$~Mpc from the source.}
  \label{fig:strain_tot}
\end{figure}

The two strain polarizations resulting from the numerical simulation are displayed in  Figs.~\ref{fig:strain_plus} and \ref{fig:strain_cross}. Among other things, one can very clearly see the time increase of the frequency of the signal close to the merging, as well as its increase in amplitude. In order to get an idea of the typical amplitude and shape of the bursts as a function of time we introduce $h_\mathrm{tot.} = \sqrt{h_+^2 + h_\times^2}$, shown in Fig.~\ref{fig:strain_tot}. The strain magnitude is irrelevant as we mostly aim at comparing with circular orbits. We recall that, even when dealing with circular orbits, the situation is tricky. Since the frequency of observation is fixed, black hole binaries with higher masses are observed closer to the merging and generate a higher strain. However this is (partially) compensated by the fact that the signal drifts faster than for smaller masses. At fixed frequency, the higher the mass, the higher the generated strain, but the shorter the time spent by the signal within the bandwidth of the detector  \cite{Franciolini:2022htd,Domcke:2023bat,barrau2023}.\\  

\begin{figure}
  \centering
  \includegraphics[width=.9\linewidth]{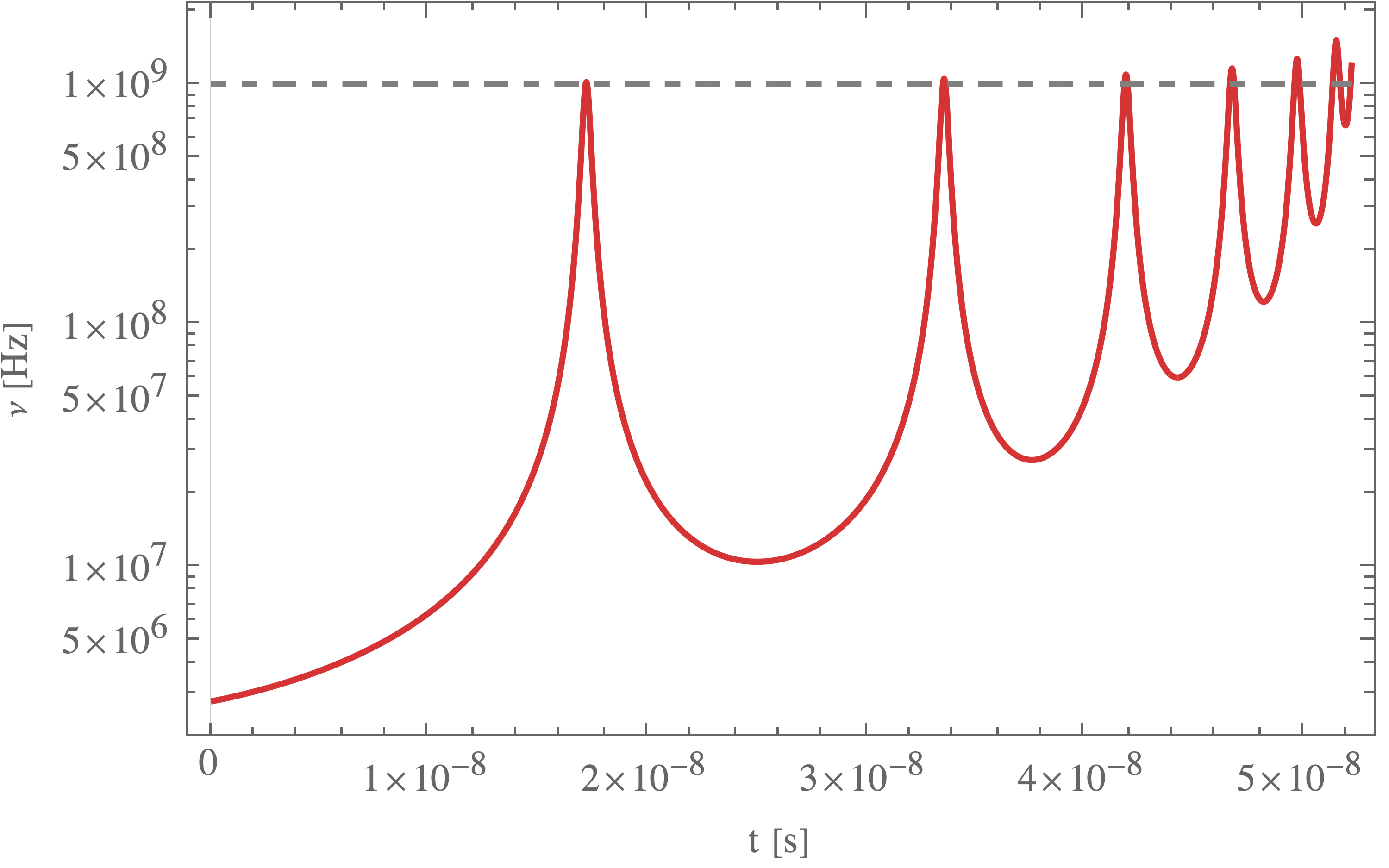}
  \caption{Time variation of the frequency. The dashed line represents the very narrow bandwidth of the detector.}
  \label{fig:freq_time}
\end{figure}

Let us now investigate the frequency of the signal. Its time evolution is displayed in  Fig.~\ref{fig:freq_time}. There are two major contributions in the evolution of this quantity. First, it can easily be noticed that the very same bursts as those appearing for the strain are also visible here. They correspond to the fact that the instantaneous angular frequency increases when the orbiting objects approach the periapsis and decreases as they get further away. Second, in addition to this fast variation, there is also an overall upward drift in the frequency which is caused by the modification of the orbital parameters $e$ and $a$ induced by backreaction. This shows that the situation can be very different than for circular orbits. In the latter case, the frequency of the signal, whose evolution is only (at the lowest order) due to backreaction, crosses the bandwidth only once whereas this can happen many times for a highly eccentric orbit. At this stage, it is far from obvious to guess which situation is the most favorable one.
Clearly, the differences become less and less pronounced as the initial eccentricity is decreased, eventually recovering the case of circular orbits for $e = 0$. The bursts  then disappear, and the two polarizations  become sinusoidal (albeit still with increasing frequency).\\ 

\begin{figure}
    \centering
    \includegraphics[width=1.0\linewidth]{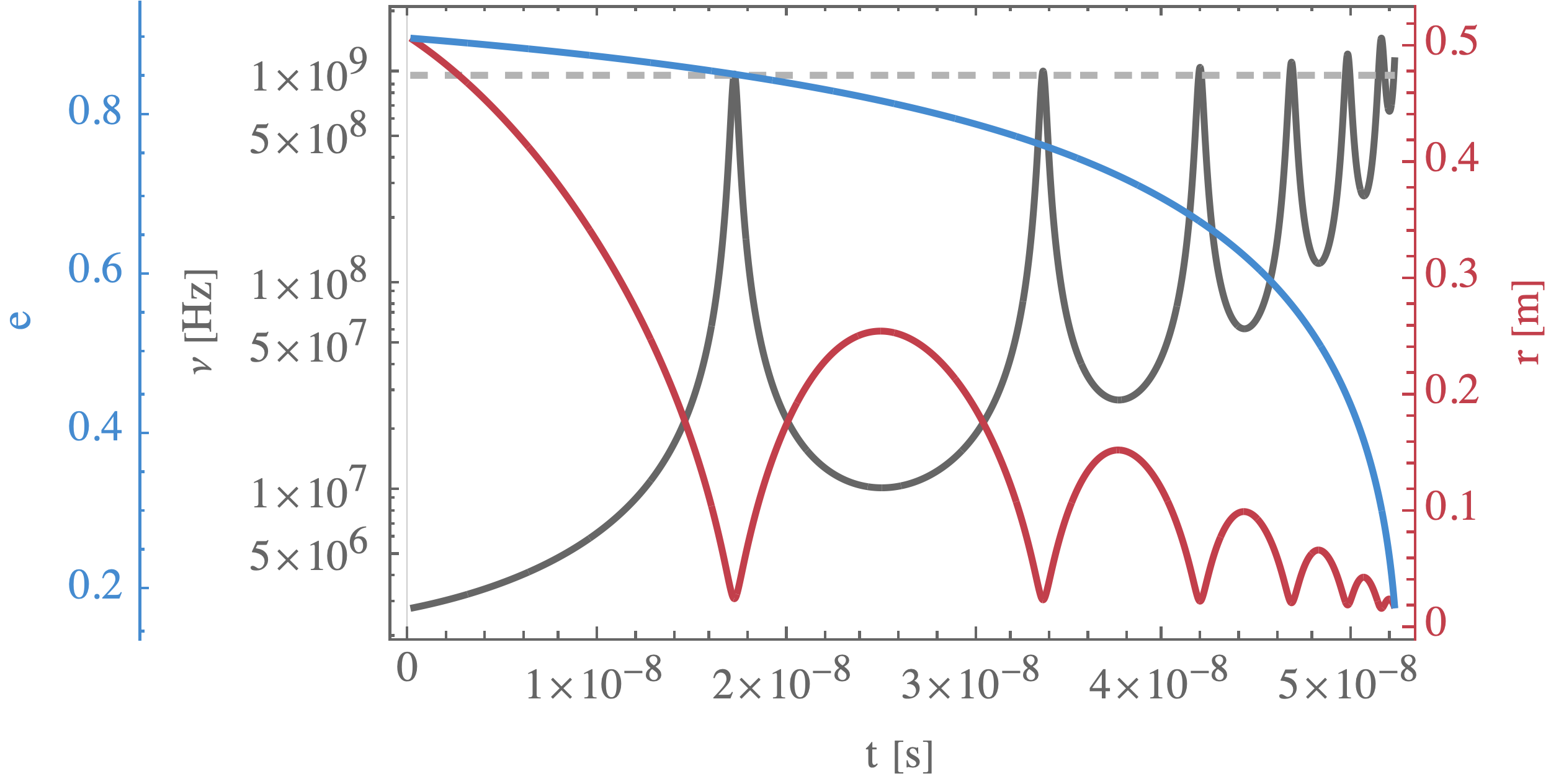}
    \caption{Time evolution of the eccentricity (in blue), orbital separation (in red), and frequency (in black), with masses $m_1 = m_2 = 1.5\times10^{-6}~M_\odot$, first eccentricity $e_1 = 0.8545$, and initial phase $\varphi_0 = -\pi$.}
    \label{fig:summary}
\end{figure}

Figure \ref{fig:summary} summarizes the situation by displaying the combined evolution of the eccentricity, orbital separation, and frequency. 

\subsection{\label{sec:3_2}Collected energy as a function of initial eccentricity and angle}

As is well known in orbital mechanics, even outside the context of gravitational waves, there is no analytical solution for the function $\varphi(t)$ except for purely circular motion. This is why a numerical integration of the equations is mandatory to get the explicit time dependence of the various physical quantities of interest. 

As we focus on detection by resonant cavities with very narrow bandwidths, the temporal characteristics of the signal are of the utmost importance, as will be made clear in the following. 

For each chosen eccentricity, we compute the time-frequency curve, similar to the one  shown in Fig.~\ref{fig:freq_time} and extract the accurate values of the times at which the frequency crosses the boundaries of the bandwidth (which, we recall, is smaller than the width of the dashed line in the plot). From these, we get the total effective duration for the signal together with the explicit time intervals over which the received gravitational power should be integrated.
Repeating this procedure for a wide range of initial conditions allows one to get a clear picture of the impact of the shape of the orbit. \\

\begin{figure}
  \centering
  \includegraphics[width=.9\linewidth]{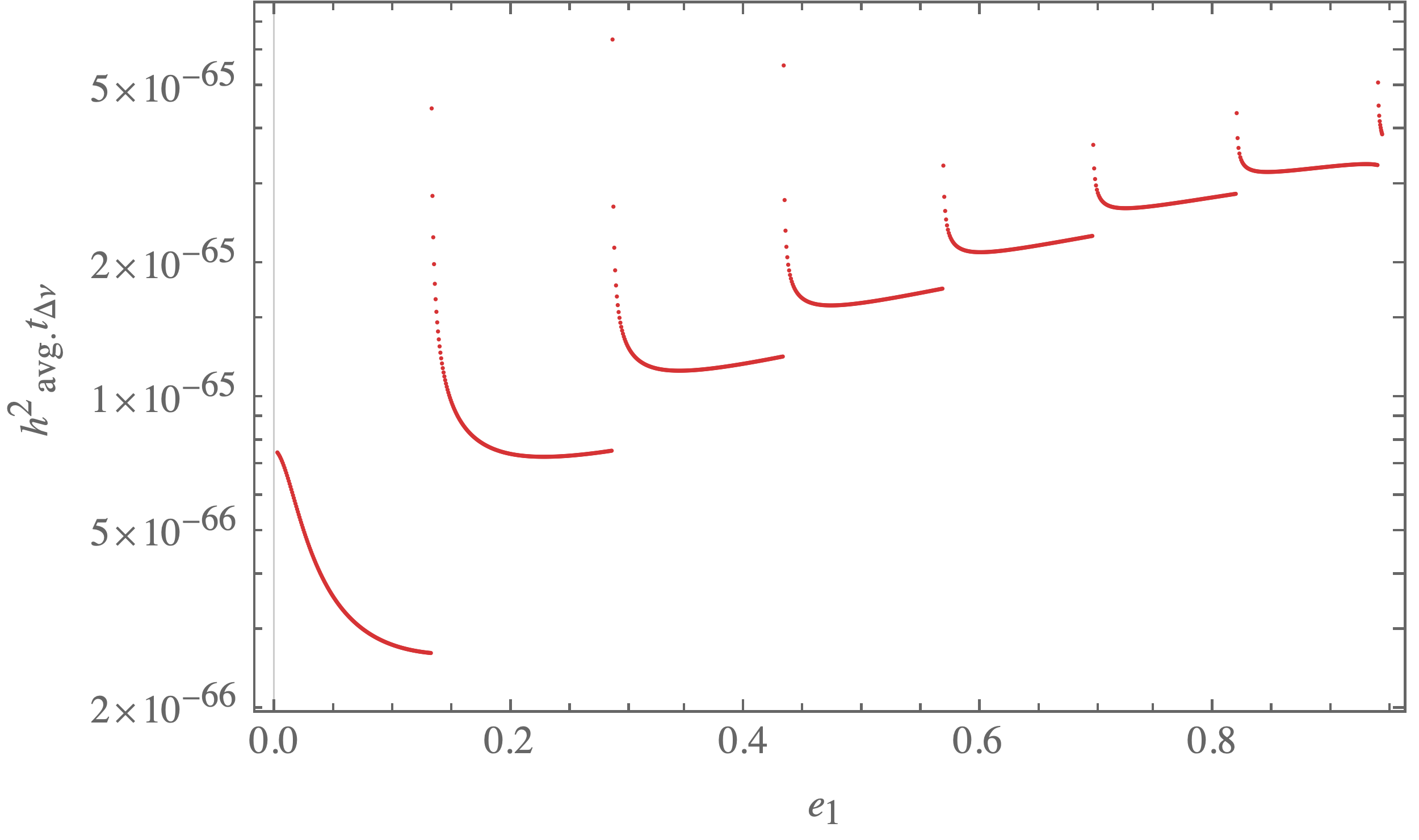}
  \caption{Amount of gravitational energy (in arbitrary units) received over the full duration of the process as a function of the (initial) eccentricity $e_1$.}
  \label{fig:energy_e}
\end{figure}

The results are displayed in Fig.~\ref{fig:energy_e} where one can distinguish three different behaviors. First, there are jumps at very specific values of the eccentricity. The physical reason for this will be made clear in the following. Second, there exists an overall trend, if the various structures are smoothed out: the received energy increases as $e_1$ increases. Finally, a clear pattern appears after each jump. 

\begin{figure}
  \centering
  \includegraphics[width=.9\linewidth]{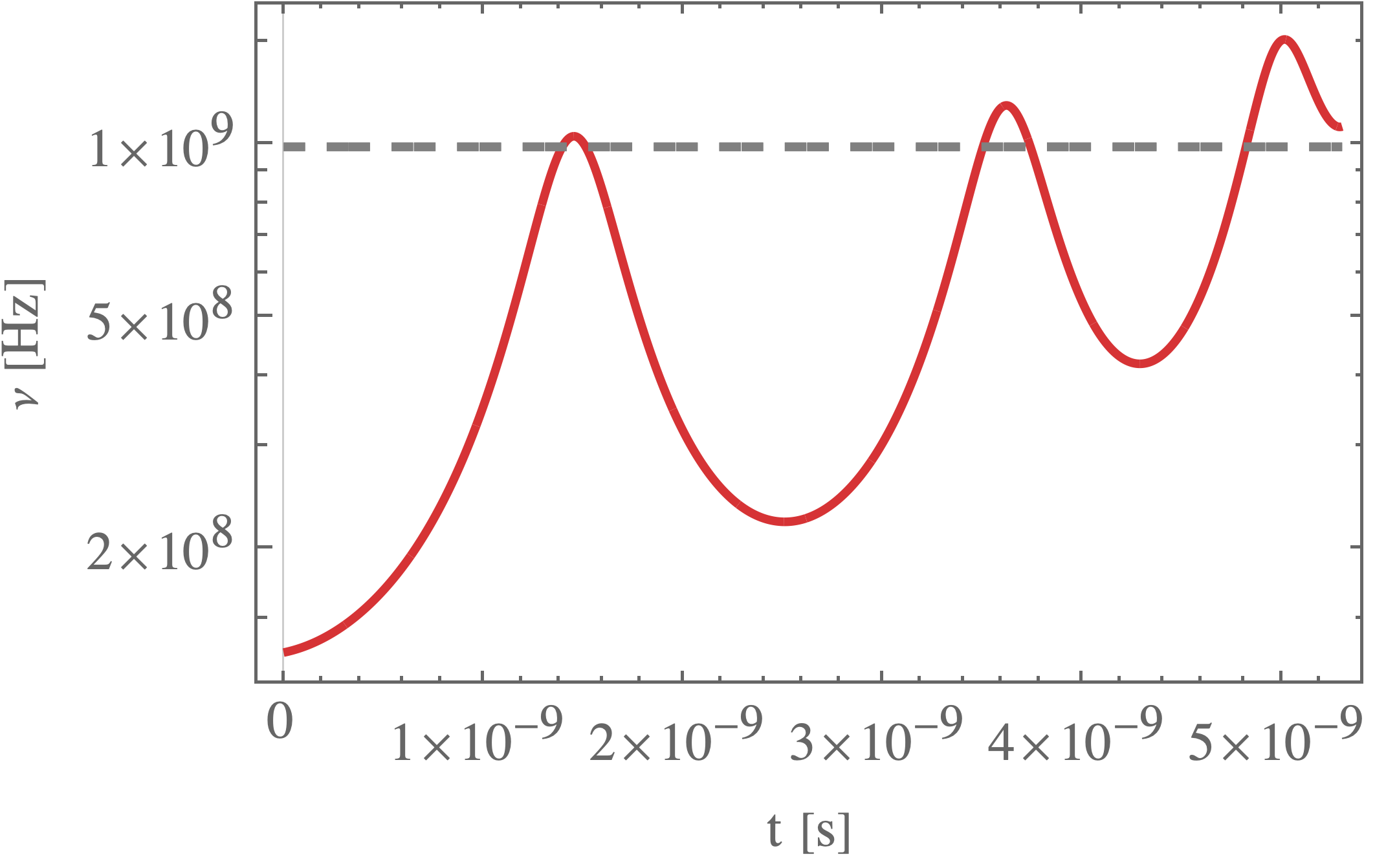}
  \caption{Time-frequency plot for a first eccentricity $e_1 = 0.423$. The horizontal line is the bandwidth of the detector.}
  \label{fig:tf_e047}
\end{figure}

\begin{figure}
  \centering
  \includegraphics[width=.9\linewidth]{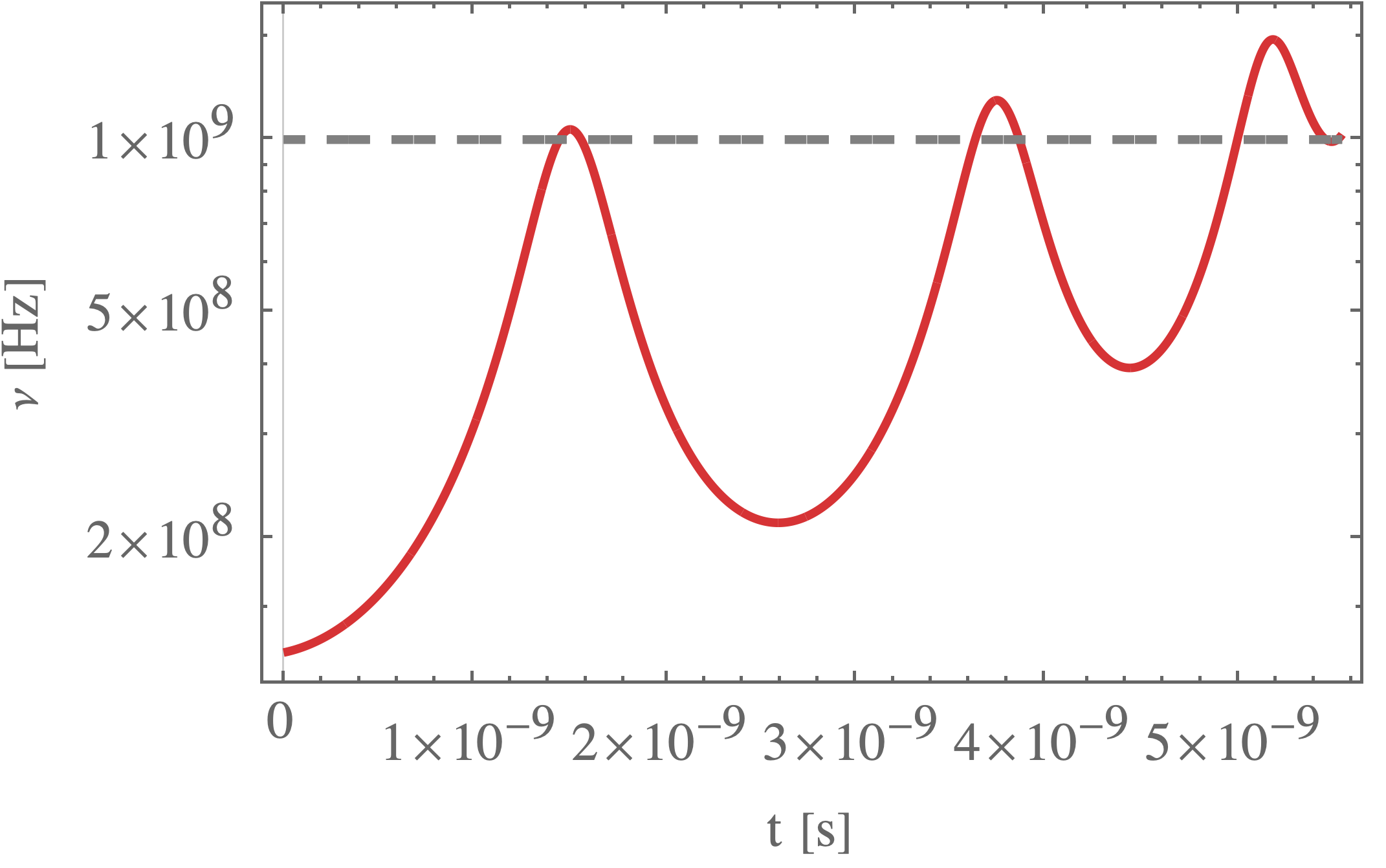}
  \caption{Time-frequency plot for a first eccentricity $e_1=0.433$. The horizontal line is the bandwidth of the detector.}
  \label{fig:tf_e048}
\end{figure}

\begin{figure}
    \centering
    \includegraphics[width=.9\linewidth]{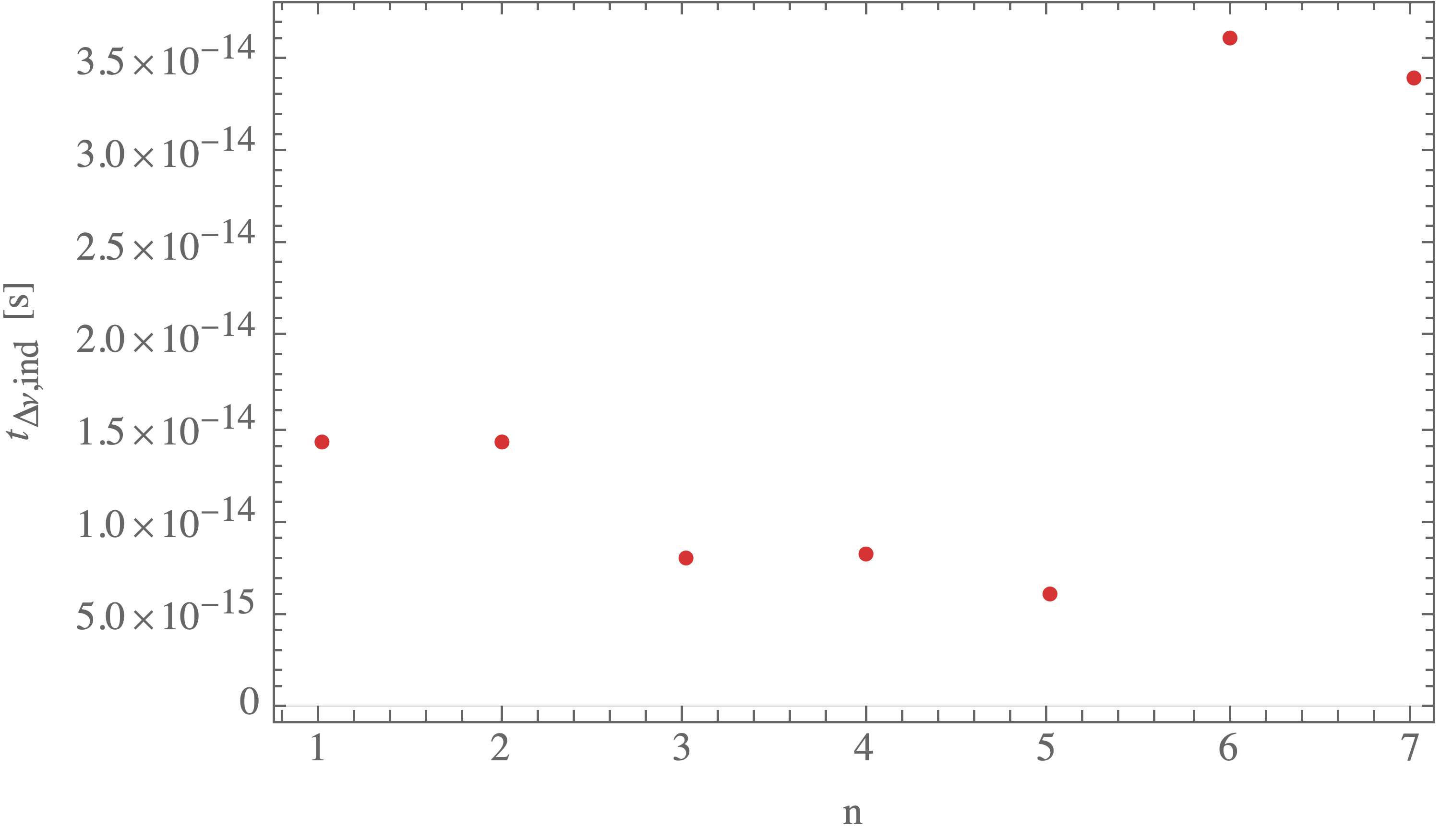}
    \caption{Time spent by the signal within the bandwidth of the detector, for each crossing, as a function of the number of crossings (between the frequency time evolution and the bandwidth). In this regime, the label $n$ happens to also be twice the number of full orbits since the beginning of the simulation.}
    \label{fig:timevslabel}
\end{figure}

In order to understand these surprising discontinuities, for instance the one around $e_1 = 0.43$, it is useful to compare the time-frequency plots corresponding to the situation just before the jump (Fig.~\ref{fig:tf_e047}) with the one corresponding to the situation just after (Fig.~\ref{fig:tf_e048}). It can clearly be seen, when comparing the two curves around $t = 5.5\times 10^{-9}$~s, that there is a bifurcation in the detected signal duration due to the fact that, in the second case, one more orbit enters the bandwidth. As it does so nearly tangentially, its contribution is very substantial, as seen from Fig.~\ref{fig:timevslabel}. We have explicitly checked that, although the strain is roughly constant at each orbit in the narrow bandwidth case considered here (except when the mass saturates the ``bound"), the time in the bandwidth can vary by several orders of magnitude. A single orbit can therefore contribute significantly more than all the others, even when the system is seen quite far from the merging.

In addition, if the eccentricity is further increased, the last minimum of the time-frequency curve is shifted downward and the curve becomes steeper. The signal hence spends a smaller amount of time in the bandwidth, thus explaining the decreasing shape of the received energy after each jump.

Finally, the average increase is mostly due to the overall dominance of the first effect over the second one. 

\subsection{\label{sec:3_3}Signal-to-noise ratio and optimal trajectories}

Let us now come to the main point of this work. Our aim is to compare elliptic trajectories with circular ones so as to determine which orbital shape maximizes the chance to detect a binary system of light black holes.\\

A first conclusion can be drawn from Fig. \ref{fig:energy_e}. The circular case, corresponding to a vanishing eccentricity, is {\it not} the best one as far as the total received energy as gravitational waves is concerned. Modulo micropatterns following the jumps, the overall trend is an increase of the energy received when the eccentricity is increased. This is already a nontrivial result as it should be kept in mind that we do not compare here, contrarily to what is often done, systems with the same initial energy, or with the same initial orbital separation, or with the same mass but without any other constraint, etc. In this study we compare systems (of fixed mass) emitting gravitational waves at the same frequency (determined by the resonant mode of the cavity). This makes this conclusion not {\it a priori} obvious.\\

In the following, we choose the mass of each object to be $5\times 10^{-7}$ solar mass, which is a smaller value than the one used up to this point. The conclusions we reach do not depend, of course, on the specific value used for the plots. The reason for decreasing the mass—hence choosing to observe the system earlier in the inspiraling process, as the frequency is fixed—is twofold. First, it allows more orbits to cross the bandwidth, which decreases the sensitivity to contingent initial conditions (that is to $\varphi_0$). The whole point of this study is to investigate the way the signal-to-noise ratio depends on the eccentricity. This question is not even correctly defined when masses are too large as the answer is, then, not univocal. It is possible to numerically search for the ``optimum" case by scanning $\varphi_0$ values. The resulting claims would, we believe, however be misleading as they would require an unreasonable amount of fine tuning, especially taking into account that the detailed structure often exhibits very narrow ``spikes." Second, smaller mass values are more realistic since, when the mass is too high—that is when the system is observed very close to the merging—the very definition of the trajectory should be revised with post-Newtonian (and post-Minkowskian) corrections that are beyond the scope of this work. Such subtleties were, anyway, not taken into account in the works on GHz signals from circular orbits to which we compare our results. The entire machinery used in this paper, as in the previous ones considering circular or hyperbolic trajectories requires one to consider masses not too close to the largest possible ones, not to mention that this corresponds to the ``generic" case as, unless one is extraordinarily lucky, there is no reason for a nearby system to have precisely the mass that maximizes the strain at the observed frequency (we remind the reader that, for circular orbits, this does not even correspond to the highest sensitivity \cite{barrau2023}).

\begin{figure}
    \centering
    \includegraphics[width=.9\linewidth]{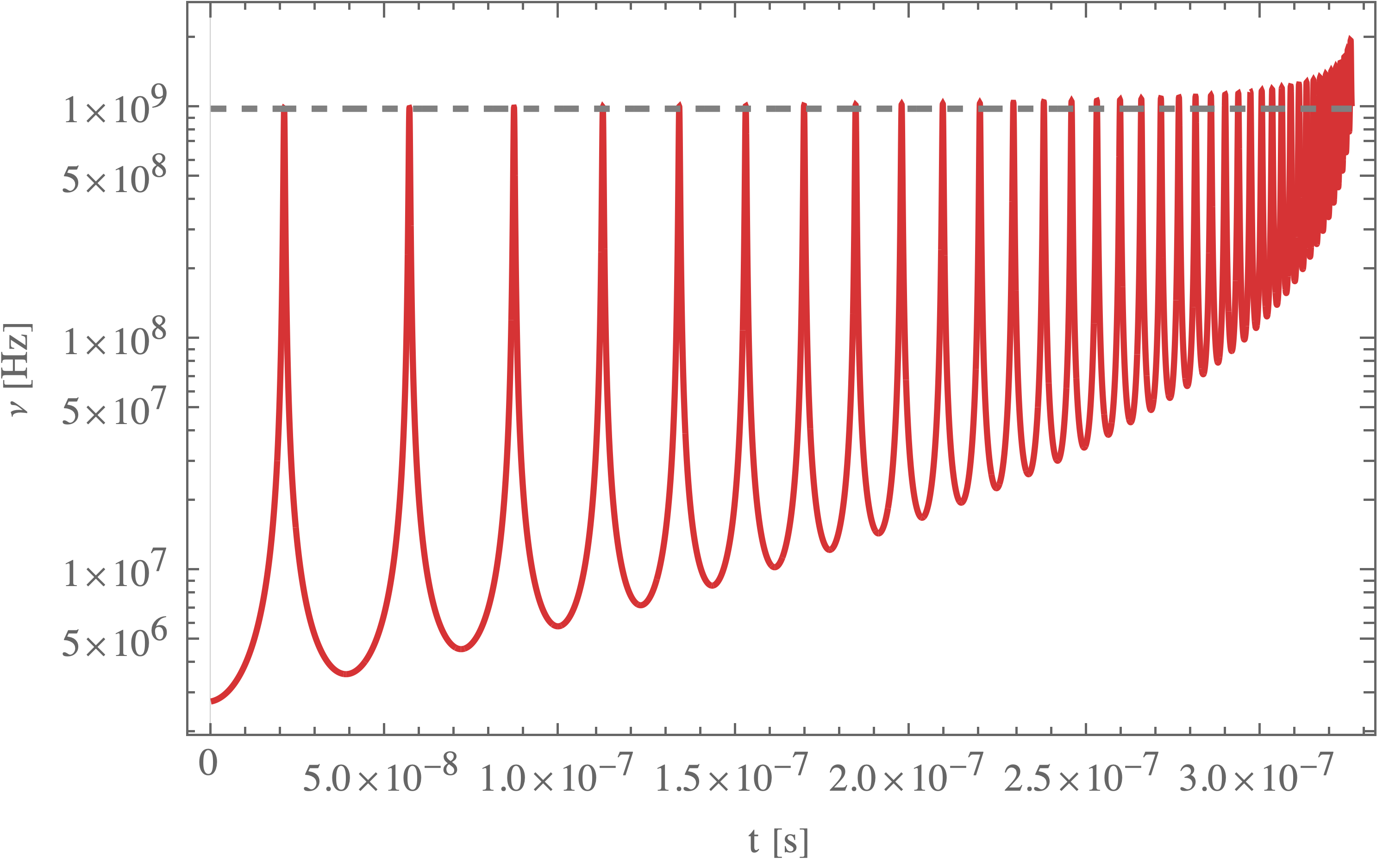}
    \caption{Time-frequency diagram for masses $m_1 = m_2 = 5\times 10^{-7}~ \textup{M}_\odot$ and first eccentricity $e_1 = 0.8931$.}
    \label{fig:smallm}
\end{figure}

A typical time-frequency diagram with $5\times 10^{-7}$ solar mass black holes is shown in Fig. \ref{fig:smallm}. It can easily be checked that the signal now crosses the bandwidth of the detector many times due to the high initial eccentricity and to the fact that the system is seen far from the merging.\\ 

The total amount of received gravitational energy is however not the final word. The Dicke  radiometer equation \cite{Sikivie:2020zpn, Berlin:2021txa, Berlin:2022hfx}, commonly used by the ``haloscopes" community to estimate the signal-to-noise ratio, reads as

\begin{equation}
\label{eq:SNR1}
    {\rm SNR} \sim \frac{P_\mathrm{sig}}{k_B T_\mathrm{sys}} \,  \sqrt{\frac{t_\mathrm{eff}}{\Delta \nu}},
 \end{equation}
 where $\Delta \nu$ is the resolution bandwidth, $T_\mathrm{sys}$ is the temperature of the system, $k_B$ is the Boltzmann constant, and $t_\mathrm{eff}$ is an effective time which will be discussed in the following (as in \cite{barrau2023}). The signal power $P_\mathrm{sig}$ is given by \cite{Berlin:2021txa}

\begin{equation}
P_{\text{sig}} = \frac{1}{2 \mu_0 c^2} Q (2\pi \nu)^3 V_\mathrm{cav}^{\frac{5}{3}} (\eta h B_0)^2,
\label{power}
\end{equation}
where $\mu_0$ is the vacuum magnetic permeability, $\nu$ the frequency of the cavity resonant mode of interest, $B_0$ the magnetic field, $Q$ the  previously discussed quality factor, $V_\mathrm{cav}$ the cavity volume, and $\eta$ a coupling coefficient set to a reasonable value of 0.1 \cite{Berlin:2021txa, Valero:2024ncz}. Once again, as we are interested in comparing the elliptic case with what happens for circular orbit, we do not need to dig into the details of most of the instrumental terms that are obviously the same for both kinds of trajectories. The important point is that the signal-to-noise ratio is proportional to $h^2Q\sqrt{t_\mathrm{eff}}$.

It is worth emphasizing that, in principle, Eq. (\ref{power}) is valid only in the steady state case. So as to keep in line with the methodology of \cite{barrau2023,barrau2023prospect}, used to investigate other trajectories, we however assume that Eq. (\ref{power}) remains a correct approximation as long at the timescale entering the SNR evaluation in Eq. (\ref{eq:SNR1}) is appropriately modified.\\

\begin{figure}
    \centering
    \includegraphics[width=.9\linewidth]{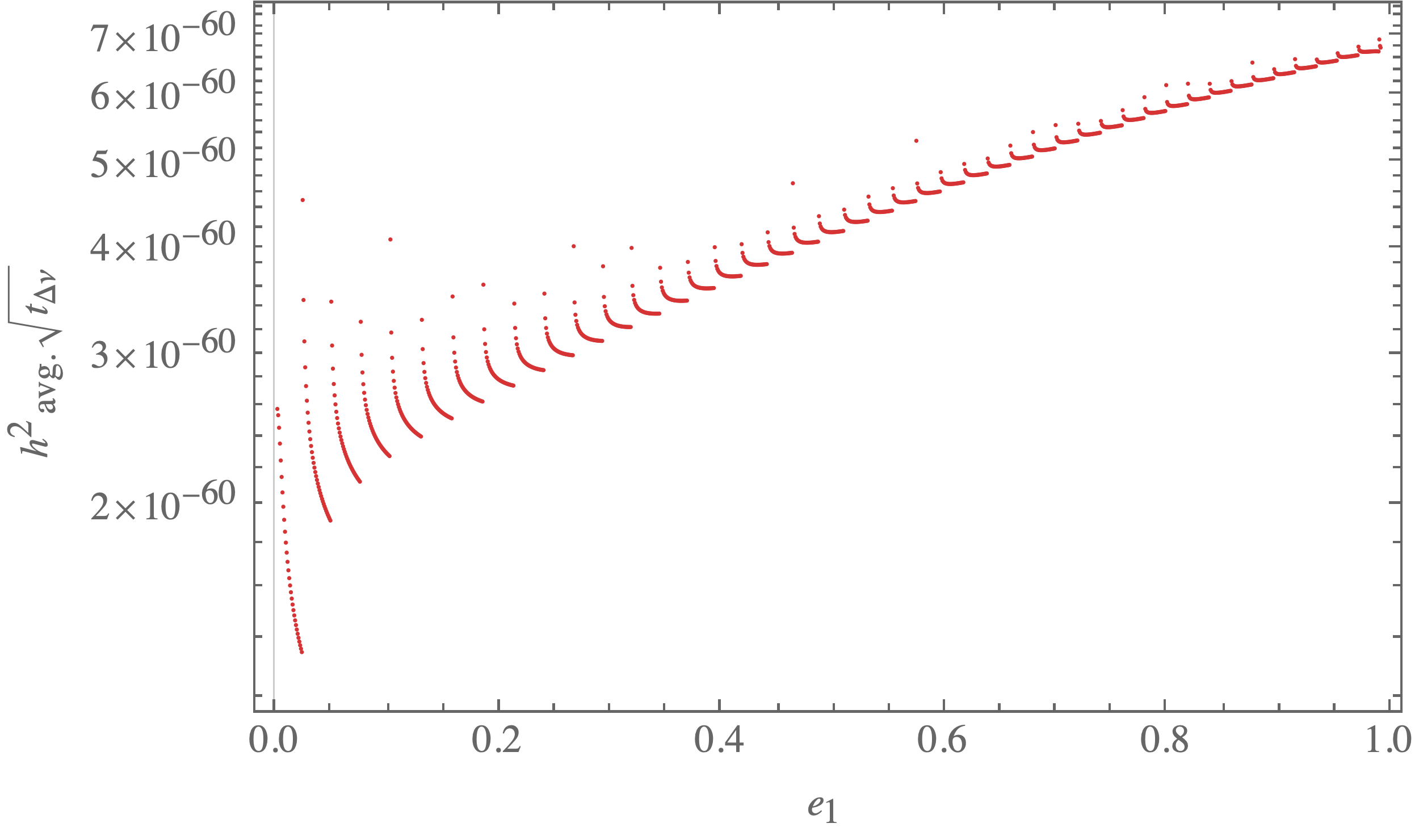}
    \caption{SNR (not normalized) as a function of the eccentricity for an effective time given by the physical time spent by the signal in the bandwidth of the detector.}
    \label{fig:teff1}
\end{figure}

The effective time $t_\mathrm{eff}$ is first assumed to be the total amount of time $t_{\Delta \nu}$ spent by the signal within the bandwidth of the detector. 
The result is given in Fig. \ref{fig:teff1}. Interestingly, modulo the expected jumps (that are now smoother as the mass has been reduced), the trend is still an increase of the sensitivity with the eccentricity. This means that in a hypothetical setting which would be purely physics limited—in the sense that the effective time appearing in the SNR would correspond to the time during which the signal frequency drifts through the detector bandwidth—highly eccentric orbits would indeed be easier to detect. It is worth stressing that, as expected, Fig. \ref{fig:teff1} is basically indistinguishable from the total energy (appropriately scaled) received as gravitational waves displayed as a function of the eccentricity.


\begin{figure}
    \centering
    \includegraphics[width=.9\linewidth]{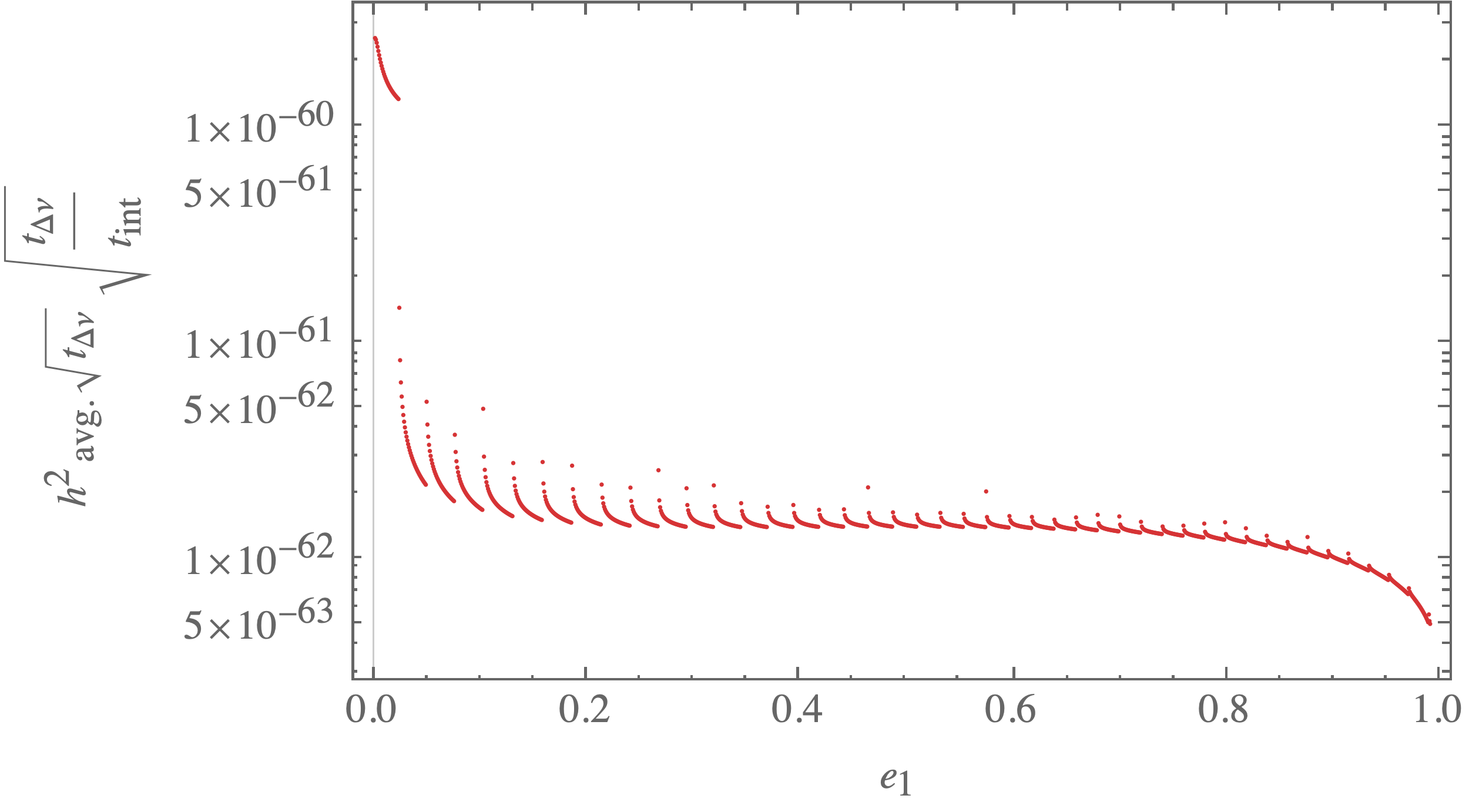}
    \caption{SNR (not normalized) as a function of the eccentricity for an effective time taking into account the competitive effect of the signal and the noise during the integration window.}
    \label{fig:teff2}
\end{figure}

This is however a fully unrealistic setting. Let us now take into account the fact the the detector will integrate the signal at least—in the very best case—between its first entrance in the bandwidth and its last exit from it. We call $t_\mathrm{int}$ this duration. The key point is that noise will also be integrated during this full window. This obliges one to modify the effective time such that, now, $t_\mathrm{eff}\sim  t_{\Delta \nu}^2/t_\mathrm{int}$. The associated results are given in Fig \ref{fig:teff2}. Very importantly, the trend is entirely reversed. The signal-to-noise ratio is now a decreasing function of the eccentricity. The circular case $e_1=0$ is now the {\it best} one. The reason is obvious: for a circular orbit, the frequency spends only one—quite long—interval of time within the bandwidth. From the viewpoint of the competition with the noise, this is clearly the best case. This effect happens to play a more important role than what is, on the other hand, gained for ellipses as an increase in $t_{\Delta \nu}$. This is the main result of this article: although the naive investigation of the received energy, presented in Fig. \ref{fig:teff1}, seems to favor highly eccentric orbits, the accurate calculation, shown in Fig. \ref{fig:teff2}, leads exactly to the opposite result. This robust conclusion does not depend on initial conditions or on free parameters.

\begin{figure}
    \centering
    \includegraphics[width=.9\linewidth]{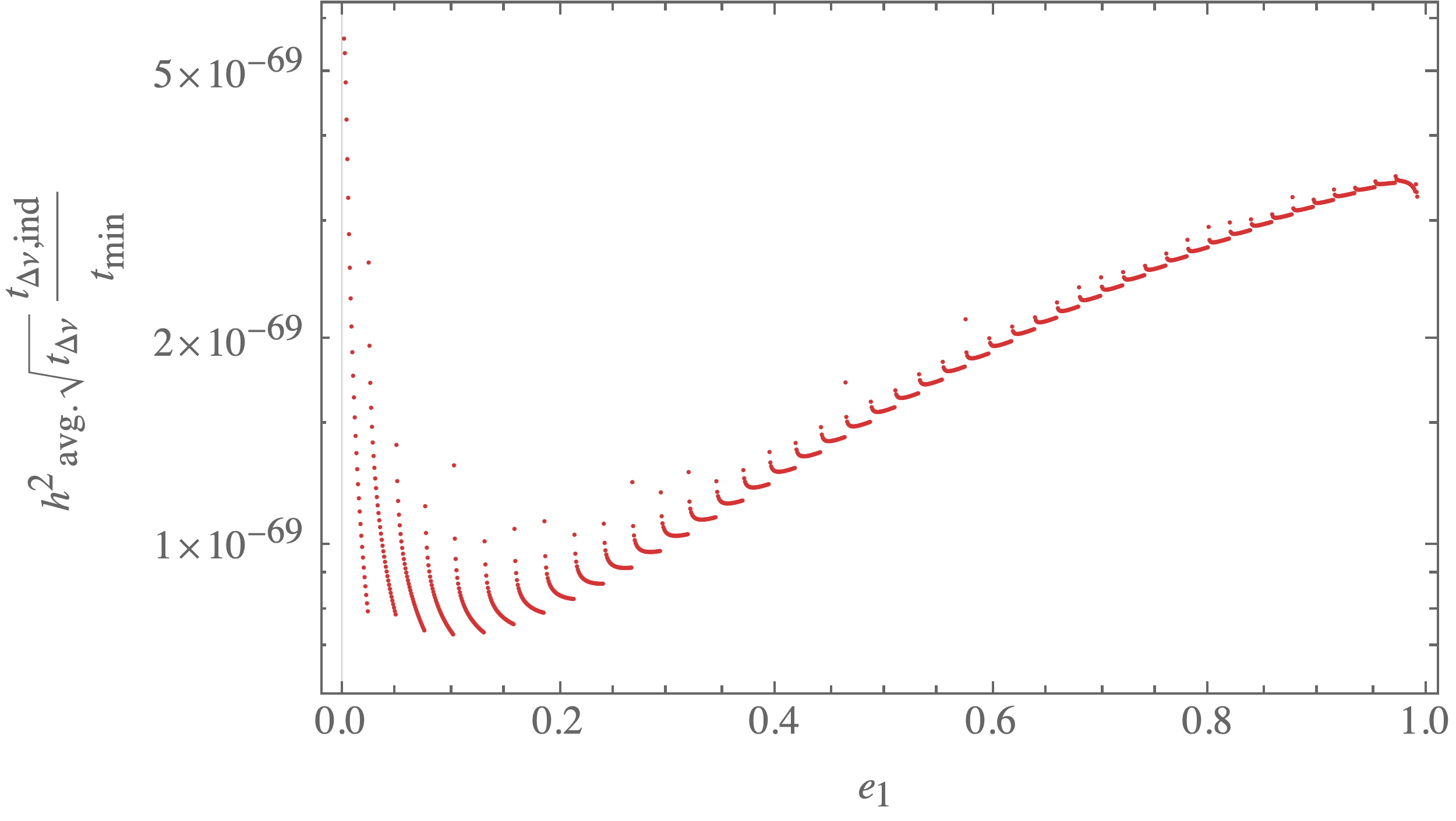}
    \caption{SNR (not normalized) as a function of the eccentricity for an effective time taking into account the charging time of the cavity.}
    \label{fig:teff3}
\end{figure}

For the sake of completeness, it is also interesting to focus on another, more subtle, effect. Owing to the finite charging time of the cavity, the quality factor entering Eq. \ref{power} should also be modified \cite{Kim:2020kfo, barrau2023}: $Q \rightarrow \nu t_{\Delta \nu} = Q t_{\Delta \nu}^\mathrm{ind} / t_\mathrm{min}$, where $t_{\Delta \nu}^\mathrm{ind}$ is the average time spent by the signal in the bandwidth for a single band crossing. This can be effectively taken into account by setting now $t_\mathrm{eff}\sim t_{\Delta \nu}\times ( t_{\Delta \nu}^\mathrm{ind}/ t_\mathrm{min})^2$. The detailed motivations for this, maybe surprising, factor are given in our previous work \cite{barrau2023}. A full simulation of the cavity response, beyond the scope of this article, is currently being developed to confirm the validity of this hypothesis. At this stage, it can be considered as a meaningfully ``worst case scenario" (therefore leading to conservative sensitivity estimates). Intuitively, this result can however be quite straightforwardly obtained by considering the convolution of the source with the cavity impulse response assumed to be a free damped oscillation. The resulting signal-to-noise ratio is presented in Fig. \ref{fig:teff3}. As expected, except for the very small values of the eccentricity, this effect is subdominant and the initial behavior is mostly recovered.

\begin{figure}
    \centering
    \includegraphics[width=.9\linewidth]{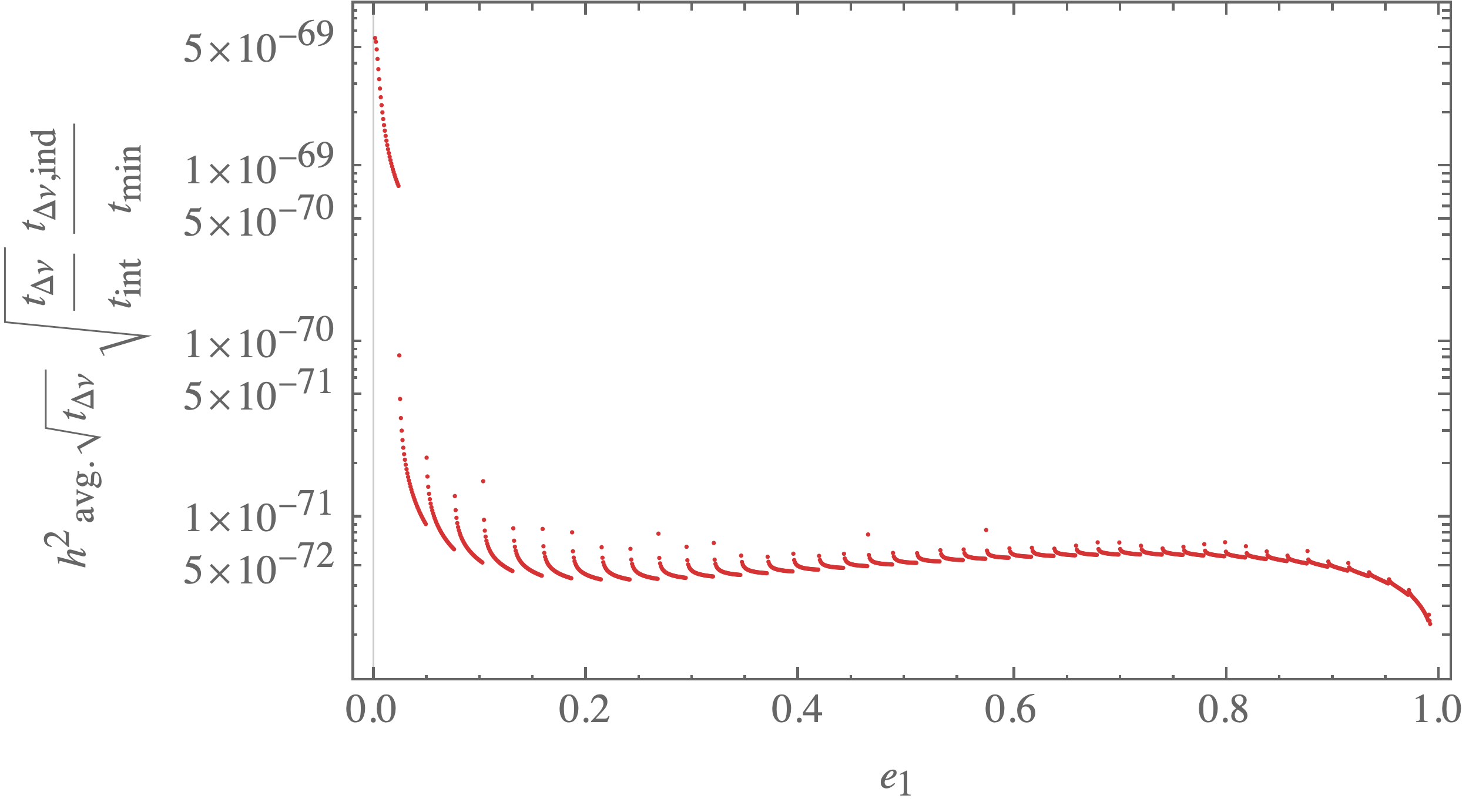}
    \caption{SNR (not normalized) as a function of the eccentricity for an effective time taking into account all the effects considered in this study.}
    \label{fig:teff4}
\end{figure}

Finally, Fig. \ref{fig:teff4} shows the evolution of the signal-to-noise ratio when all effects are taken into account. Clearly, the circular orbit ($e_1=0$) is the one leading to the best situation. Quite interestingly, the overall shape is however nearly a ``plateau" between $e_1=0.1$ and $e_1=0.9$. The signal-to-noise ratio is just strongly boosted for very small eccentricities and strongly damped for very high ones.

\begin{figure}
    \centering
    \includegraphics[width=.9\linewidth]{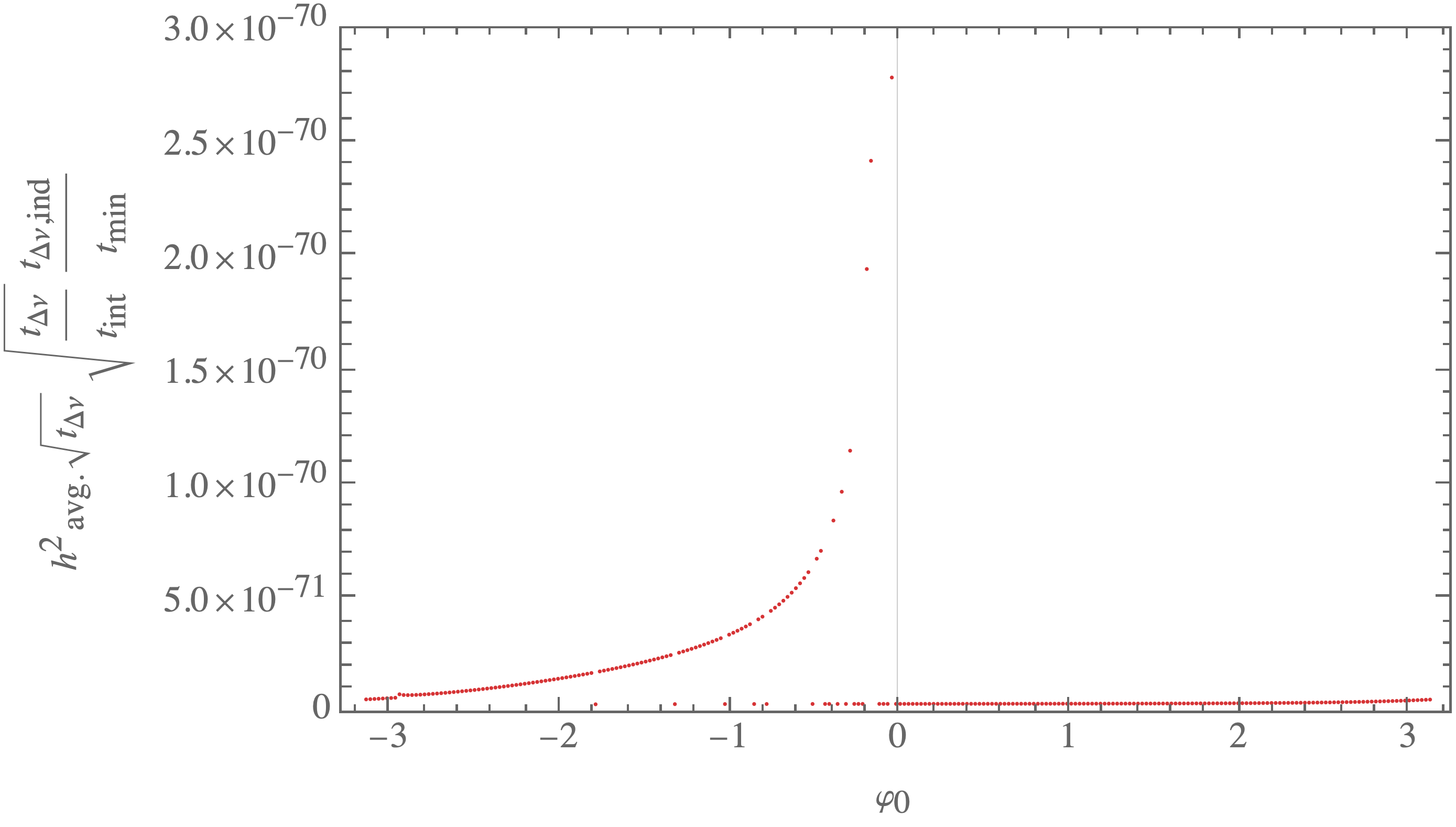}
    \caption{SNR (not normalized) as a function of the initial phase $\varphi_0$ with initial eccentricity $e_0 = 0.6$ for an effective time taking into account all the effects considered in this study.}
    \label{fig:teff4_phi}
\end{figure}

\begin{figure}
    \centering
    \includegraphics[width=.9\linewidth]{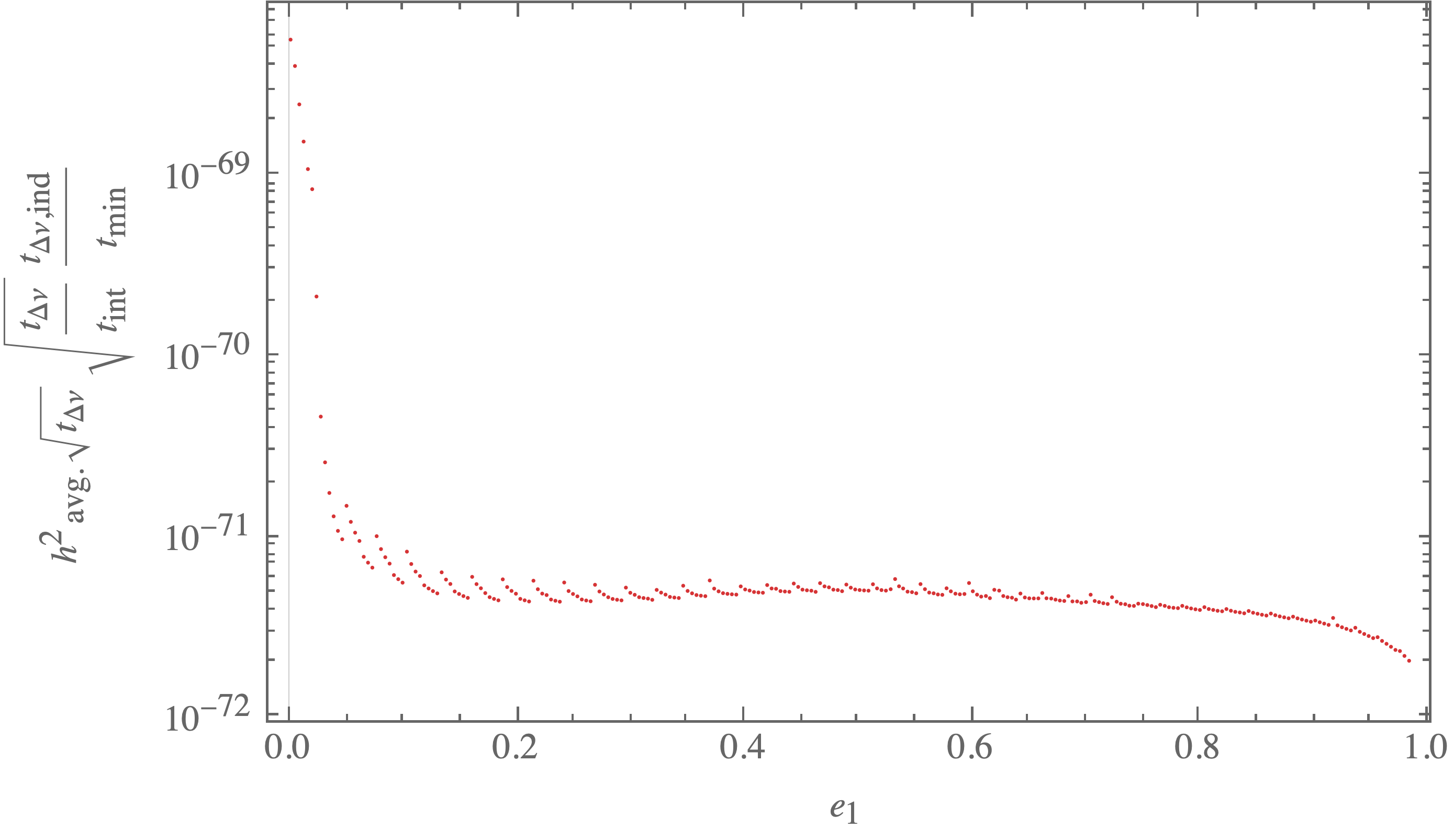}
    \caption{SNR (not normalized) as a function of the eccentricity for an effective time taking into account all the effects considered in this study, averaged over the initial phase $\varphi_0$.}
    \label{fig:teff4_med}
\end{figure}

To be as exhaustive as possible, we have considered two variations on Fig.~\ref{fig:teff4}. First, in Fig.~\ref{fig:teff4_phi} the SNR is investigated as a function of the initial phase $\phi_0$, as discussed previously (it should be underlined that one extreme point, around $10^{-69}$ has been left out of the plot in order to focus on the main structure of the $\varphi_0$ dependence). Importantly, the points lying significantly above the baseline are, actually, not as relevant as one might have guessed at first glance. This is for two reasons. First, it should be kept in mind than many points with very low SNR are actually interspersed between high SNR points—exhibiting the slightly chaotic nature of the system where, we recall, a slight change in initial conditions can lead to a departure from the \textit{resonance} displayed in Figs.~\ref{fig:tf_e047} and \ref{fig:tf_e048}. This means that these optimal trajectories are actually very improbable when one considers a reasonable range of conditions. A second, more quantitative, argument—which is related to the first one—is the following: if one takes the average (respectively median) SNR over the full range of $\varphi_0$ values, the resulting value is $2.4\times 10^{-71}$ (respectively $4.7\times 10^{-72}$), which confirms the statement that the result obtained with $\varphi_0 = -\pi$ is much closer to being typical than the few isolated extrema that can be found by varying the initial phase. This makes sense and this is well motivated if one comes back to the physical meaning of this work: as observers on Earth, we have no knowledge of the specific parameters of the system possibly observed; rather we have to measure what Nature gives us which—as we show here—will be in general in a very suboptimal configuration. 
While it is true that the initial phase does obviously change the SNR, it is a very slight dependence for all practical purposes. 
If, however, one insists on taking this into account, it is possible to average over $\varphi_0$. This adds quite a lot of numerical complexity for a small correction with no consequence on the conclusions. Still, we display the result of this heavy procedure in Fig.~\ref{fig:teff4_med}.

\begin{figure}
    \centering
    \includegraphics[width=.9\linewidth]{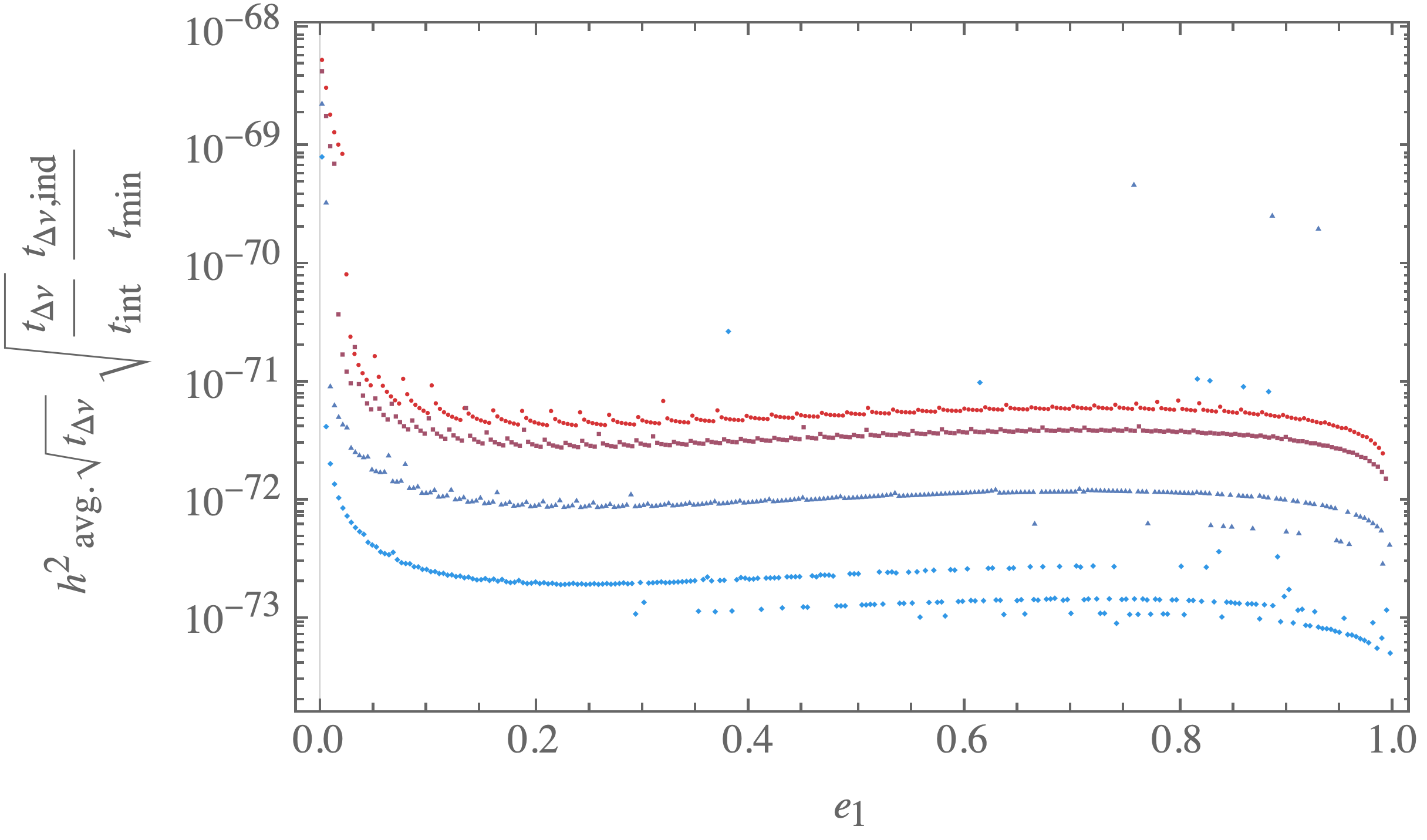}
    \caption{SNR (not normalized) as a function of the eccentricity for an effective time taking into account all the effects considered in this study,for a total mass $m_1 + m_2 = 10^{-6}~M_\odot$ and different mass asymmetries $\frac{m_1 - m_2}{m_1 + m_2}$. From top to bottom : $0$, $0.6$, $0.9$, $0.98$.}
    \label{fig:teff4_asym}
\end{figure}

Finally, in Fig.~\ref{fig:teff4_asym} the masses $m_1$ and $m_2$ have been varied while keeping a constant total mass $M = m_1 + m_2$, \textit{i.e.}, we have introduced an asymmetry $\frac{m_1 - m_2}{m_1 + m_2}$. As expected, the SNR decreases as the asymmetry increases. Although not surprising, it is once again not as obvious as the case of elliptic orbits. All quantities of interest are expressed with a prefactor written as a positive power of the reduced mass which, since the total mass is constant, is itself proportional only to $m_1 m_2 = m_1(M - m_1)$. The obvious maximum is reached at $M/2$, which means that the mass asymmetry should only decrease the SNR. This argument is however rigorous only for circular orbits as the full factorization does not hold anymore in the case of elliptic ones where the intricate time dependence of the orbital quantities could, in principle, introduce subtle effects. This plot however shows that this has no practical consequence.

\section{\label{sec:4}Discussion and conclusion}

This work does not aim at giving a definitive answer to the raised question but tries to clarify the situation at the lowest nontrivial order. The conclusion reached, in this framework, is clear and reliable. Several approximations were however made and should be explicitly listed.\\

First, as previously stated, the (textbook) time averaging procedure used is not {\it a priori} fully satisfactory. In principle, it is possible to solve the dynamics: the strain depends on the matrix elements $M_{ij}$ which, themselves, depend on  $\varphi$, $\omega_p$, and $e$. All those variables are coupled and time dependent which makes the numerical resolution lengthy and rather unstable. We have however explicitly checked that our conclusions are unchanged when the time averaging procedure is replaced by a full integration. 

Second, we have implicitly assumed—and this is related with the previous point—that the trajectory remains elliptic ``at each time step," with parameters evolving smoothly.  This is precisely what is also being done in studies on GHz gravitational waves from circular orbits to which those results are compared. The textbook formulas used for the time evolution of the instantaneous frequency assume a circle at each instant. This is an approximation which breaks down at the end of the process even when ignoring post-Newtonian corrections: very close to the merging, the trajectory is no longer quasicircular (or quasielliptic). Calculating the strain in this regime is of course a well-known and widely discussed question. It remains mostly irrelevant for this work. We focus on masses well below the upper bound imposed by the frequency, the latter corresponding to a system observed at the merging.

Third, post-Newtonian corrections could, in principle, be taken into account (see, {\it e.g.} \cite{Levi:2018nxp} and references therein). We insist that the main conclusion being, at this stage, that the signal is deeply out of reach, the need for including subtle relativistic effects is not currently crucial. Orders of magnitude first need to be known so that the haloscope community understands whether it is worth, or not, trying to optimize resonant cavities for this quest.

Fourth, the signal was assumed to be monochromatic at each instant with a pulsation given by $\dot{\varphi}$. This corresponds to the peak of the Fourier transform, which is obviously an approximation. 

Fifth, the Dick radiometer formula used to calculate the power left by the gravitational wave in the cavity is certainly not the final word on this question. In this work, we have refined its use as much as we could—as it drastically impacts the results—but more refined estimates could be used in the future. Our steps are, intentionally, exactly the same as in the previous works to which we confront our results. Whatever the possible refinements that might be considered in the future, they would affect all trajectories in the same way, and it is extremely unlikely that the conclusion would be reversed.\\

In spite of all these restrictions, our main results are clear. It was expected that elliptic trajectories might, thanks to the bursts they generate, improve the sensitivity estimates for very high frequency gravitational waves from compact binary systems. We have shown that the total power received by the cavity is indeed larger than for circular trajectories. However, when taking into account the complicated time structure of the signal and its consequence on the measurement performed, the conclusion is fully reversed: the higher the eccentricity, the lower the signal-to-noise ratio. \\

We conclude that the upper limit on the distance at which a binary system of black holes can be detected—derived {\it e.g.}, in \cite{barrau2023}—can only be decreased when considering highly eccentric trajectories. No detection is therefore to be expected with this technique in the near future. 
Of course, should the bandwidth be very different, or another analysis technique be used ({\it e.g.}, based on the temporal aspect of the signal), higher eccentricities could become interesting to consider as we have shown that the total amount of energy available is still higher than for circular orbits. This is particularly relevant since the Newtonian framework used here (with its clear limits) is not limited to the range of masses we have considered—the equations and results derived in this work would also apply to higher masses and lower signal frequencies.\\

We emphasize that, beyond the nontrivial conclusion that was reached, favoring circular orbits for current setups, the subtleties of elliptic orbits combined with narrow-band detection deserved clarification, even at the Newtonian order.

\section*{Acknowledgement}

We thank Juan Garc\'ia-Bellido for suggesting this study.

\section*{Data availability}
No data were created or analyzed in this study.

\section{Appendix}

\begin{figure}
  \centering
  \includegraphics[width=.8\linewidth]{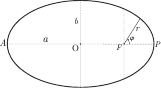}
  \caption{Main parameters of an elliptic orbit. The center is $O$ and the focus is $F$ while $A$ and $P$ are respectively the apoapsis and periapsis—the furthest and closest points from the focus. The semimajor and semiminor axes are $a$ and $b$ whereas $r$ and $\varphi$ are the coordinates of the moving object.}
  \label{fig:ellipse}
\end{figure}

Figure~\ref{fig:ellipse} shows the parametrization used for elliptic trajectories. \\



The leading order derivation of the strain generated is  textbook \cite{maggiore2008}. For simplicity, we assume the detector to be far away from the objects and along an observation direction perpendicular to the orbital plane.
The two strain polarizations in the transverse-traceless gauge at a distance $R$ from the source are given by
\begin{align}
  h_+ &= \frac{G}{R c^4}(\ddot{M}_{11} - \ddot{M}_{22}),\label{eq:pol_plus}\\
  h_\times &= \frac{G}{R c^4}(2\ddot{M}_{12}),\label{eq:pol_cross}
\end{align}
with $M_{ij}$ the second mass moment related to the quadrupole moment $Q_{i j}$ by
\begin{equation}
  Q_{ij} = M_{ij} - \frac{1}{3}M_k^k \delta_{ij}.
\end{equation}
Written in matrix form, it reads as
\begin{equation}
  M_{ij} = \mu r^2(\varphi)\begin{pmatrix}
    \cos^2\varphi & \cos\varphi\sin\varphi & 0\\
    \cos\varphi\sin\varphi & \sin^2\varphi & 0\\
    0 & 0 & 0
  \end{pmatrix}.
\end{equation}

Some care must be taken at this point concerning the time derivatives of this object, as well as some time integrals which will appear in the next section. Not only do the coordinates $\varphi$ and $r$ depend on time, but the orbital parameters $\omega_p$ and $e$, which appear in the function $r(\varphi)$, are also time dependent. This adds another layer of subtlety in the calculation that is often overlooked. One usually considers, in deriving the expressions for the strain, that only the dynamical variables related to the radiating objects, $\varphi$ and $r$, do vary and assume that the orbital parameters themselves are constant, reintroducing their time dependence later on through some averaging procedure. As shown in this article, this is indeed sufficient for this study although this is not {\it a priori} obvious.
Under this hypothesis, one is led to the  expressions for the strains given by Eqs. (\ref{eq:hplus}) and (\ref{eq:hcross}).

The radiated energy and angular momentum are \cite{maggiore2008}
\begin{align}
  \dv{E}{t} &= \frac{2G}{15c^5}\bqty{\dddot{M}_{11}^2 + \dddot{M}_{22}^2 + 3 \dddot{M}_{12}^2 - \dddot{M}_{11} \dddot{M}_{22}},\\
  \dv{L}{t} &= \frac{4 G}{5c^5}\bqty{\ddot{M}_{12}\pqty{\dddot{M}_{11}-\dddot{M}_{22}}}.
\end{align}

Following the usual procedure, 
we define the period-averaged derivatives:
\begin{align}
  \dot{E}_{\mathrm{avg}} &= \frac{1}{T}\int_0^T\dd{t}\dv{E}{t},\\
  \dot{L}_{\mathrm{avg}} &= \frac{1}{T}\int_0^T\dd{t}\dv{L}{t}.
\end{align}
Using these equations and the expression for the second mass moment $M_{ij}$, one straightforwardly obtains the two radiated quantities in terms of the eccentricity $e$ and semimajor axis $a$:
\begin{align}
  \dot{E}_{\mathrm{avg}} &= -\frac{\mu^2 G}{15c^5}\frac{\kappa^3}{a^5}\frac{1}{(1-e^2)^\frac{7}{2}}\pqty{96 + 292e^2 + 37e^4},\\
  \dot{L}_{\mathrm{avg}} &= -\frac{\mu^2 G}{15c^5}\frac{\kappa^\frac{5}{2}}{a^\frac{7}{2}}\frac{1}{(1-e^2)^2}\pqty{96 + 84e^2}.
\end{align}

It is more convenient to work with the semimajor axis $a$ instead of the angular velocity at the periapsis $\omega_p$ as the equations are much simpler and intuitive that way. 
The total energy and angular momentum then read as
\begin{align}
  E &= -\frac{\kappa\mu}{2a},\\
  L &= \mu \sqrt{\kappa a \pqty{1 - e^2}}.
\end{align}
In the case of open conics the semimajor axis $a$ becomes infinite (for the parabola) or negative (for hyperbolae), and the energy becomes, as expected, positive. Differentiating these last two equations with respect to time, and combining them with the previously given results for $\dot{E}_{\mathrm{avg}}$ and $\dot{L}_{\mathrm{avg}}$, one is led to the system Eqs. (\ref{eq:adiff}) and (\ref{eq:ediff}).\\

For the sake of completeness, it is worth recalling that the frequency spectrum of the radiated power can also be computed for a Keplerian elliptic orbit. Performing the calculation in Fourier space and using the quadrupole formula, one is led, for the $n$-th harmonic, to \cite{maggiore2008}
\begin{equation}
    P_n=\frac{32G^4\mu^2M^3}{5c^2a^5}g(n,e),
\end{equation}
with
\begin{equation}
    g(n,e)=\frac{n^6}{96a^4}\left( A_n^2(e)+B_n^2(e)+3C_n^2(e)-A_n(e)B_n(e) \right),
\end{equation}
the coefficients being given by a combination of Bessel functions:
\begin{equation}
    A_n=\frac{a^2}{n}\left( J_{n-2}(ne)-J_{n+2}(ne)-2e(J_{n-1}(ne)-J_{n+1}(ne)) \right),
\end{equation}
\begin{equation}
    B_n=\frac{b^2}{n}\left( J_{n+2}(ne)-J_{n+2}(ne)\right),
\end{equation}
\begin{equation}
    C_n=\frac{ab}{n}\left( J_{n-2}(ne)+J_{n+2}(ne)-e(J_{n+1}(ne)+J_{n-1}(ne)) \right).
\end{equation}

It is basically a monotonically decreasing function of $n$ in the limit $e\rightarrow 0$ while it peaks on higher harmonics as the trajectory becomes more and more eccentric.

\bibliography{elliptic}

\end{document}